\documentclass{aastex631}

\usepackage{amsmath}
\usepackage{subfigure}
\usepackage{natbib}
\usepackage{todonotes}
\usepackage{booktabs}

\usepackage{hyperref}
\hypersetup{
    colorlinks=true,
    linkcolor=blue,
    filecolor=magenta,      
    urlcolor=blue,
    pdftitle={Overleaf Example},
    pdfpagemode=FullScreen,
    }

\usepackage{dcolumn}
\usepackage{bm}

\usepackage{orcidlink}

\usepackage{graphicx}

\begin{document}

\title{Multi-Tracer Correlated Stacking: A Novel Way to Discover Anisotropy in nano-Hz Stochastic Gravitational Wave Background}

\author{Mohit Raj Sah \orcidlink{0009-0005-9881-1788}}\email{mohit.sah@tifr.res.in}
\author{Suvodip Mukherjee \orcidlink{0000-0002-3373-5236}}\email{suvodip@tifr.res.in}
\affiliation{Department of Astronomy and Astrophysics, Tata Institute of Fundamental Research, Mumbai 400005, India}

\begin{abstract}
The isotropic stochastic gravitational wave background (SGWB) generated by a population of supermassive black hole binaries (SMBHBs) provides a unique window into their cosmic evolution. In addition to the isotropic power spectrum, the anisotropic component of the signal carries additional information about the supermassive black holes (SMBHs) and host galaxy connection. The measurement of this signal is usually carried out by angular power spectra, which is only a sufficient measure for a Gaussian and statistically isotropic distribution of SMBHBs, where the statistical properties of a field remain unchanged across the sky. In contrast, the contribution from  SMBHBs in nano-hertz SGWB will be hosted by fewer massive galaxies, making the nano-hertz background anisotropic and non-Gaussian. As a result, the performance of angular power spectra in extracting the underlying physics is limited. In this work, we propose a novel technique called the \texttt{Multi-Tracer Correlated Stacking}, which enables the detection of anisotropies in the SGWB by stacking the signal from regions of the sky with tracers of BHs such as active galactic nucleus (AGNs), quasars, bright galaxies, etc., that can be mapped up to high redshift.  We demonstrate this technique on a simulated supermassive BHBs distribution using an AGN catalog, which maps the underlying matter distribution approximately up to redshift $z=5$. This stacking technique uniquely distinguishes between isotropic and anisotropic distributions of SGWB source, surpassing the capabilities of angular power spectrum-based methods in detecting anisotropic signals. This highlights the effectiveness of this technique in detecting anisotropic SGWB signals and in the future, this technique can play a crucial role in its discovery.
\end{abstract}


\section{Introduction} \label{sec:intro}
The detection of gravitational waves (GWs) from the black hole binary (BHB) merger GW150914 by the LIGO-Virgo Collaboration \citep{abbott2016observation} marked the beginning of GW astronomy. Additionally, the recent evidence of a stochastic gravitational wave background (SGWB) at nanohertz (nHz) frequencies \citep{agazie2023nanograv,antoniadis2023second,zic2023parkes,xu2023searching} has opened new avenues for exploring the universe. At nHz frequencies, the most probable source of the SGWB is the population of inspiraling supermassive BHBs \citep{sesana2008stochastic,sesana2013gravitational,burke2019astrophysics}. As these binaries inspiral and emit GWs, they collectively produce low-frequency SGWB detectable by pulsar timing arrays (PTAs) \citep{sesana2013gravitational,burke2019astrophysics}. 

While the isotropic component of the SGWB has been the primary focus of initial detections, increasing attention is now being directed toward the potential anisotropies in the background \citep{taylor2013searching, mingarelli2013characterizing,taylor2020bright,sato2024exploring,gardiner2024beyond,sah2024imprints}. Recent searches for anisotropies in datasets, including the NANOGrav 15-year data, have found no significant evidence, instead setting upper limits on the possible level of anisotropy in the GWB \citep{agazie2023bnanograv}. These anisotropies can arise due to the spatial clustering of the sources, which can be detected using cross-correlation with galaxies as demonstrated for the first time by \citet{Sah:2024etc}, and can be useful for studying the cosmic evolution of supermassive BHs \citep{sah2024imprints,Semenzato:2024mtn}. 
 In addition to clustering, Poissonian fluctuation and variations in the properties of individual binaries, such as mass, orbital separation, etc, can further contribute to fluctuations in the GW signal across the sky  \citep{saeedzadeh2023shining,sato2024exploring,sah2024imprints}. Detecting these anisotropies would provide a complementary perspective to the isotropic GW background, offering a unique opportunity to study the cosmic distribution and evolution of supermassive BHs \citep{gardiner2024beyond,sah2024imprints,lemke2024detecting,Sah:2024etc,semenzato2024cross}.

\begin{figure}
    \centering    \includegraphics[width=16cm]{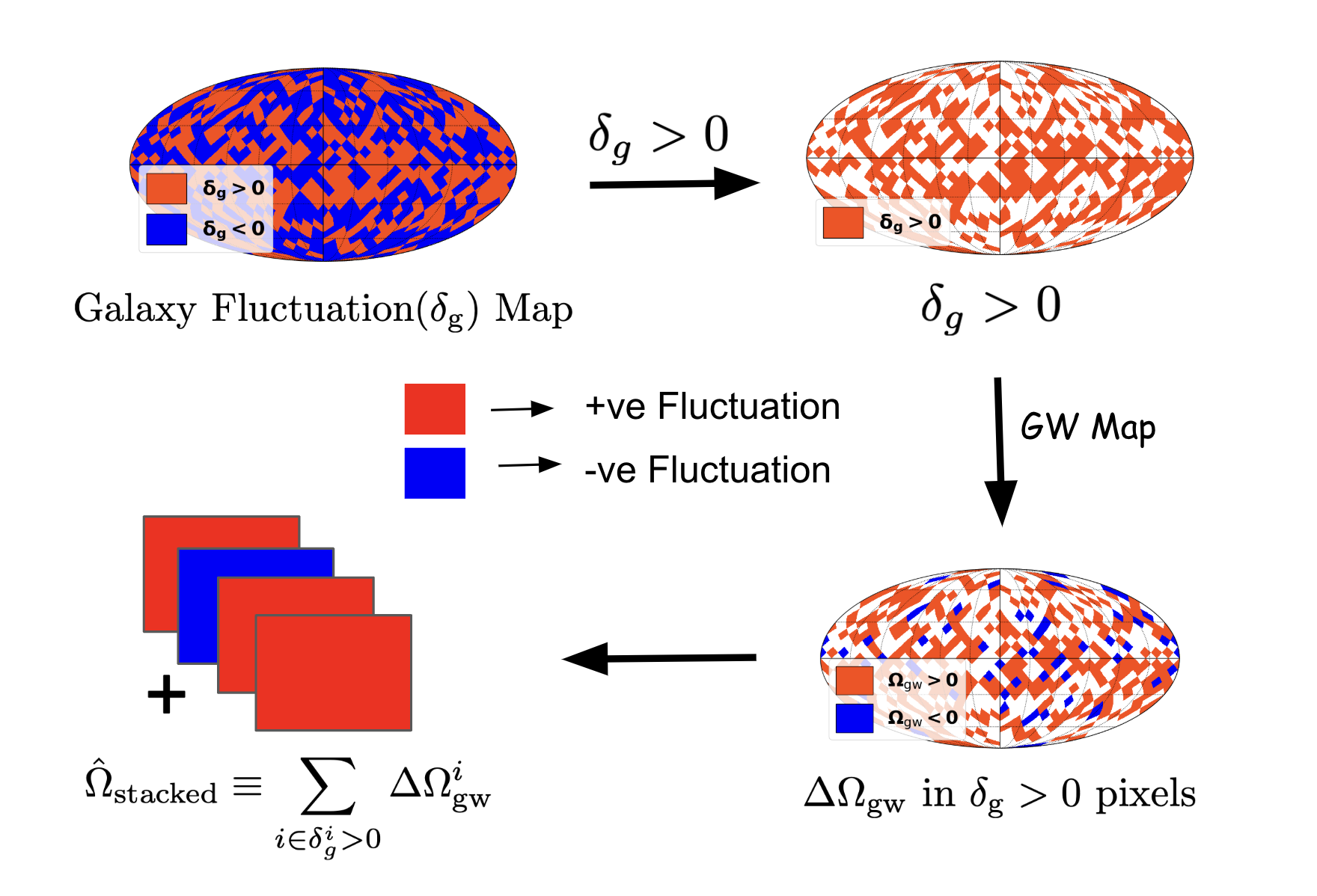}
    \caption{Schematic diagram demonstrating the stacking technique. 
    The pixels with positive fluctuation in the galaxy number density ($\delta_g$) are identified. The stacked signal is obtained by summing the SGWB density fluctuation ($\Delta\Omega_{\rm GW}$) in these pixels. Specifically, for each selected pixel $i$ with $\delta_g(i) > 0$, the corresponding $\Delta\Omega^{i}_{\rm GW}$ values are summed over all such pixels. 
    If the GW sources follow the galaxy distribution, the stacked signal will show a large positive value.}
    \label{fig:Map}
\end{figure}

In this paper, we present a novel technique, \texttt{Multi-Tracer Correlated Stacking}, designed to detect non-Gaussian anisotropies in the nHz SGWB by stacking the signal from pixels that are rich in galaxies. The idea is based on the understanding that supermassive BHBs are hosted by massive galaxies; therefore, the SGWB generated by the supermassive BHB population is expected to trace the distribution of these galaxies. This technique capitalizes on the multi-tracer approach, where the galaxy distribution serves as a tracer for the underlying SGWB sources. Fig. \ref{fig:Map} illustrates the schematic diagram summarizing this technique. The pixels with a positive fluctuation in galaxy number density $(\delta_g)$ are identified. For each selected pixel $i$ with $\delta^{i}_g > 0$, the corresponding fluctuation in the SGWB density, $\Delta\Omega^{i}_{\rm GW}$, is summed over all such pixels to obtain the stacked signal. The high galaxy number density pixels are identified using a catalog observed from large-scale structure surveys. Since astrophysical GW sources, such as supermassive BHBs, are expected to reside in galaxies, these high-density pixels should correspond to regions of enhanced SGWB. Consequently, summing the SGWB fluctuations $(\Delta\Omega_{\rm GW})$ from these pixels should yield a large positive stacked signal.

By correlating the galaxy distribution with the SGWB signal and stacking contributions from pixels with galaxy overdensities, this method not only detects the imprint of anisotropy but also establishes a strong correlation between different tracers of the BHBs and the SGWB sources. Additionally, it provides a deep insight into which types of galaxies are more likely to host SMBHBs, which is critical for identifying the nature and origin of the SGWB in the nHz frequency range.

The structure of the paper is as follows: Section \ref{Comp} provides a comparison between the effectiveness of the \texttt{Multi-Tracer Correlated Stacking} technique and the conventional power spectrum analysis; in Sec. \ref{Sim_pop}, we detail the population models of supermassive BHBs and the simulations used to generate the SGWB based on these models; in Sec. \ref{Form}, we describe the formalism of the \texttt{Multi-Tracer Correlated Stacking} technique;
Sec. \ref{App}, demonstrates the application of the stacking technique on the AGN catalog, showcasing its performance under various scenarios. Finally, Sec. \ref{Conc} concludes the paper with a discussion of key findings and future prospects.

Here, $\delta^{i}_g$ and $\Delta\Omega^{i}_{\rm gw}$ represent the overdensities of galaxies and SGWB density, respectively, in pixel $i$.

\begin{figure}[ht]
  \centering
  \subfigure[]{\label{fig:dl_2}
    \includegraphics[width=0.8\textwidth]{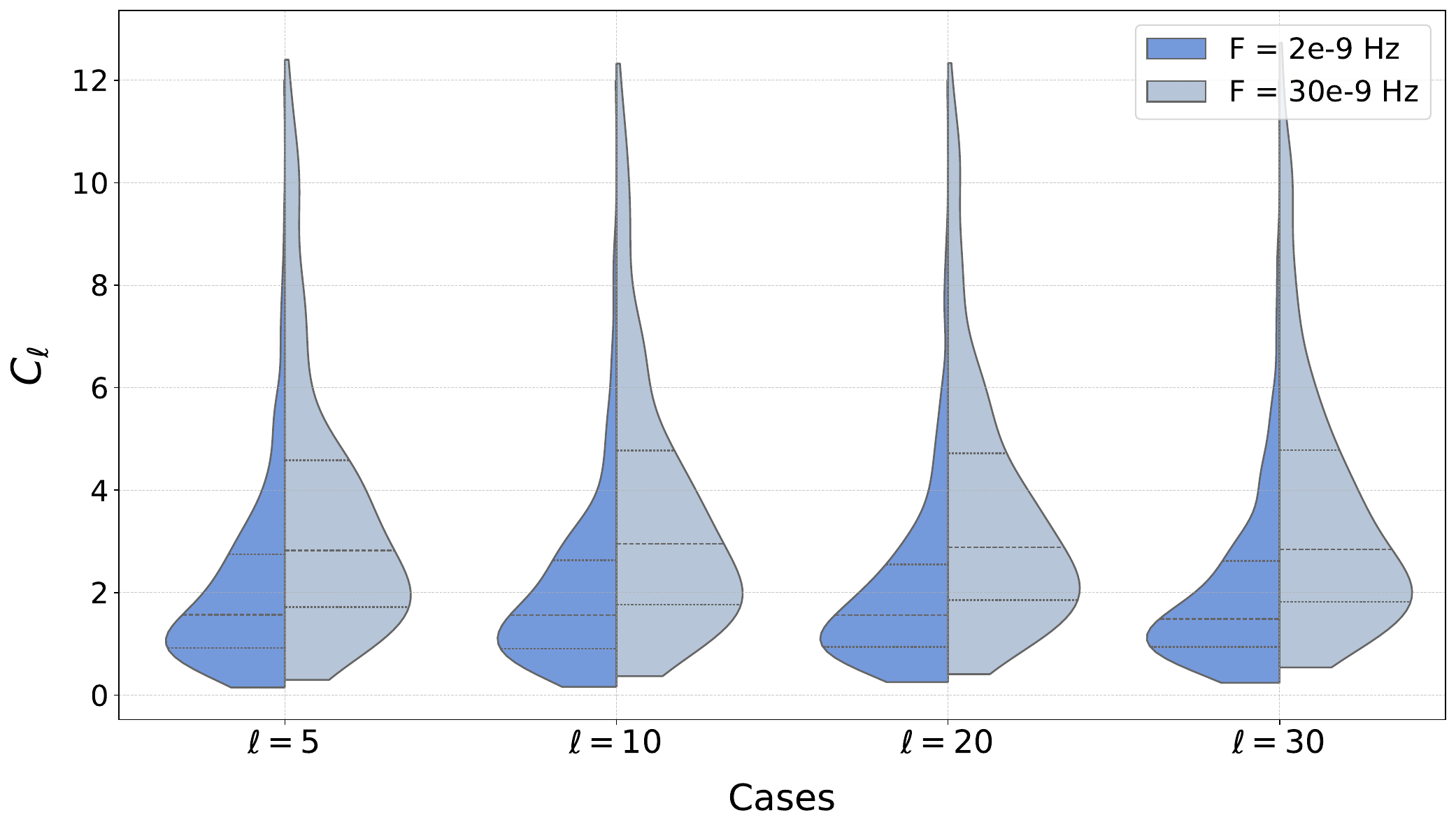}}
  \hspace{0.02\textwidth} 
  \subfigure[]{\label{fig:m1_2}
    \includegraphics[width=0.8\textwidth]{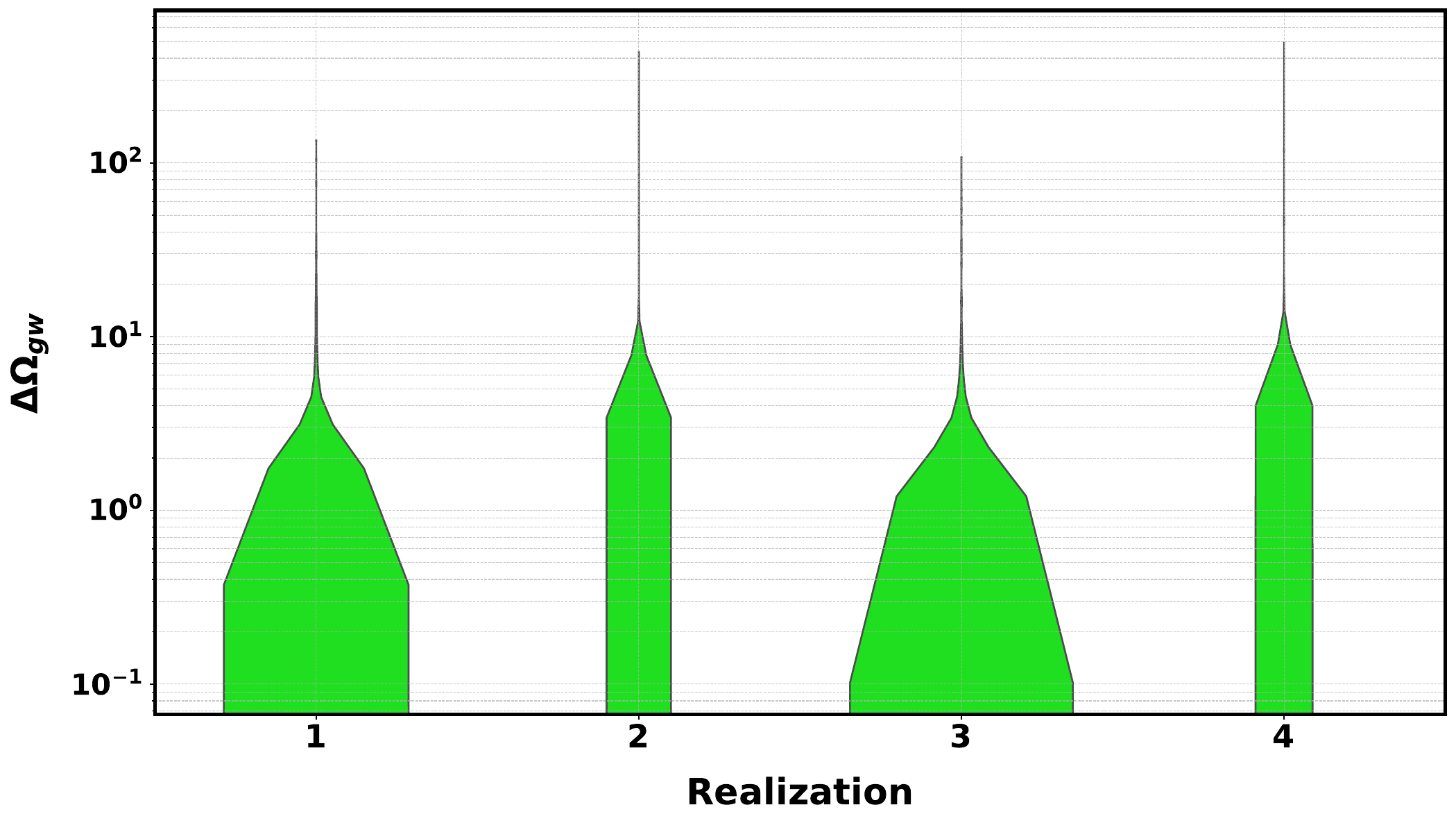}}
  \caption{(a) Violin plot showing the distribution of $C_\ell$ values across different realizations of the supermassive BH population. (b) Distribution of the fluctuations in $\Omega_{\rm gw}(\hat{n})$ across the sky for four different realizations.}
  \label{Hist_Cl}
\end{figure}

\begin{figure}
    \centering    \includegraphics[width=18cm]{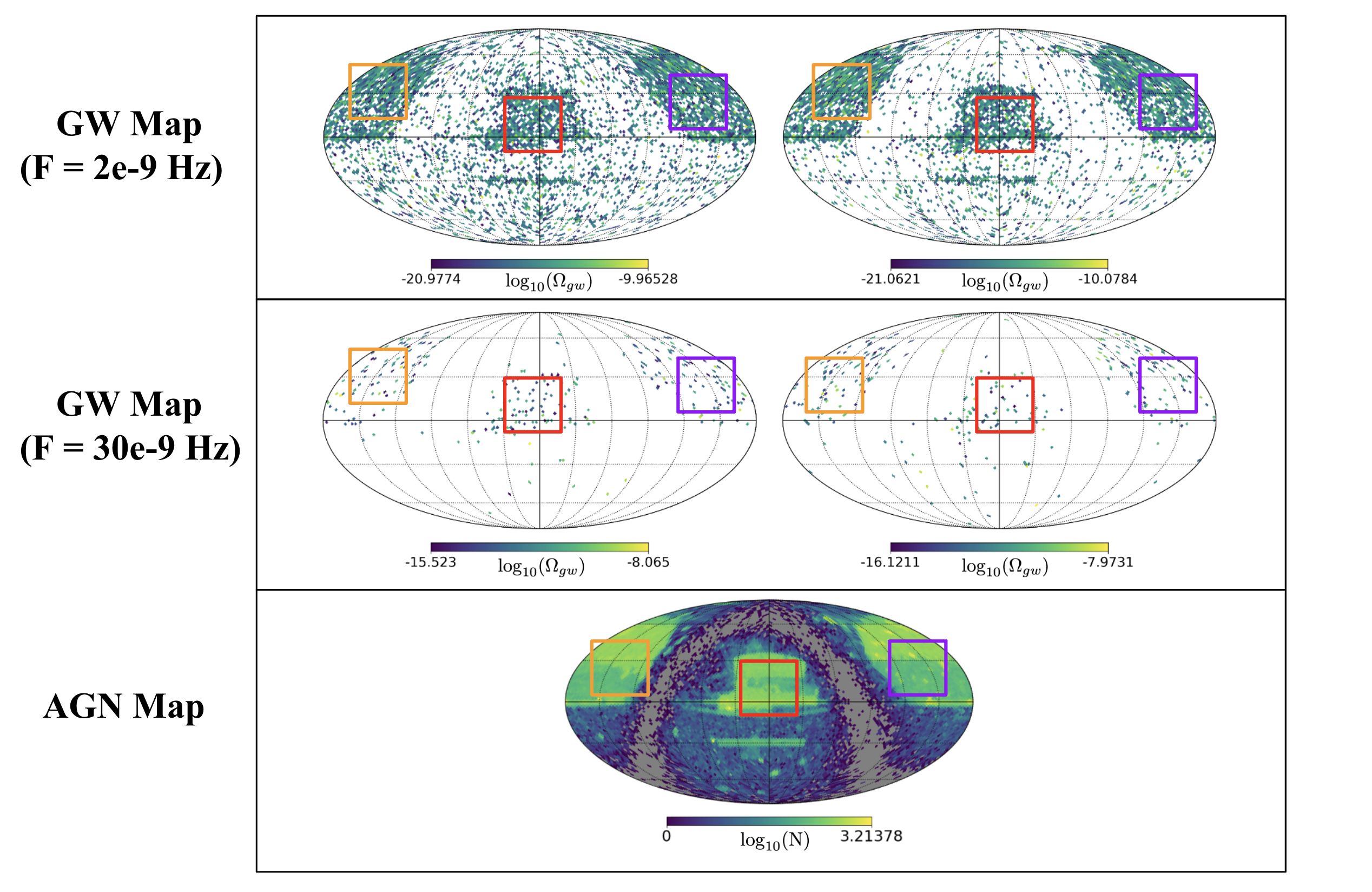}
    \caption{SGWB maps at two different frequencies and for two different realizations of the GW source population. The AGN catalog map, shown alongside, serves as a tracer for the underlying supermassive BH distribution. Colored square boxes highlight regions of high galaxy density in the AGN map and their corresponding regions in the SGWB maps.}
    \label{fig:Map1}
\end{figure}

 \begin{figure*}
    \centering
    \begin{minipage}{0.4\linewidth}
        \centering
        \subfigure[]{\label{M1_Ms_1}
        \includegraphics[width=\linewidth]{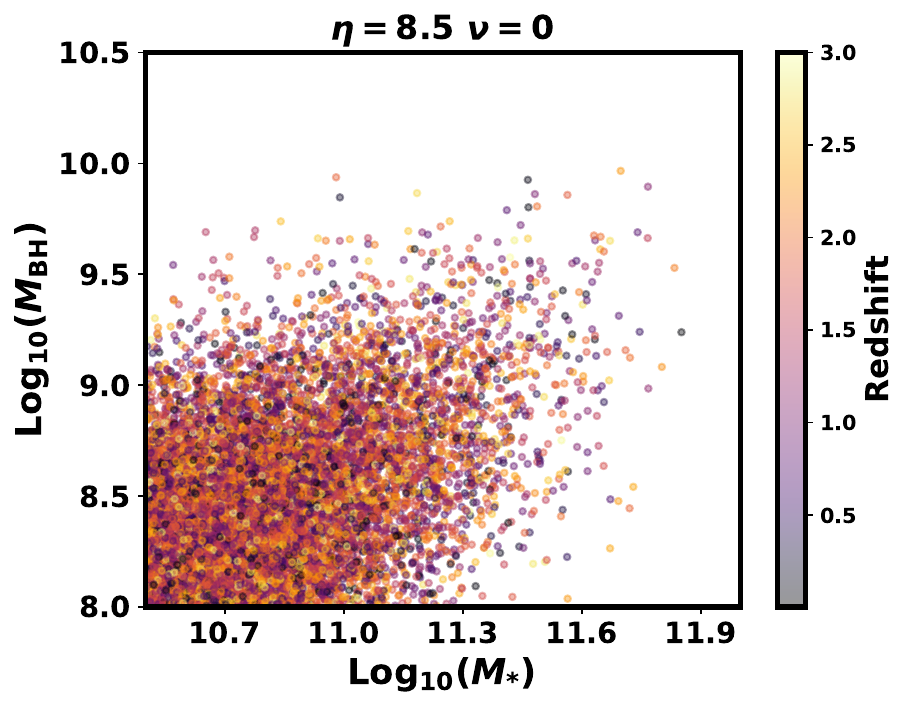}}
    \end{minipage}%
    \begin{minipage}{0.4\linewidth}
        \centering
        \subfigure[]{\label{M1_Ms_2}
        \includegraphics[width=\linewidth]{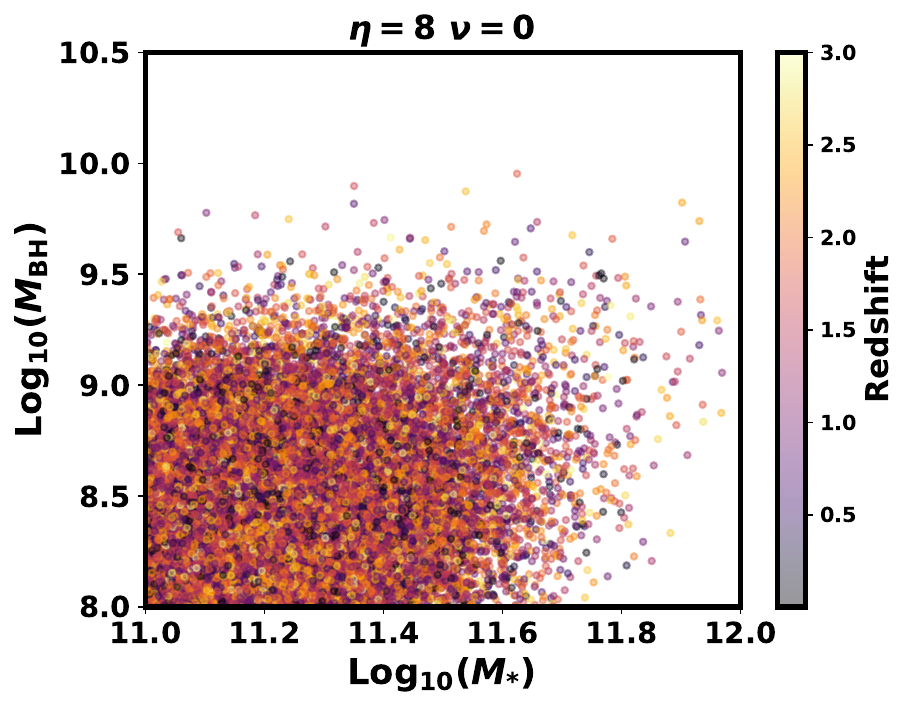}}
    \end{minipage}%
    \begin{minipage}{0.4\linewidth}
        \centering
        \subfigure[]{\label{M1_Ms_3}
        \includegraphics[width=\linewidth]{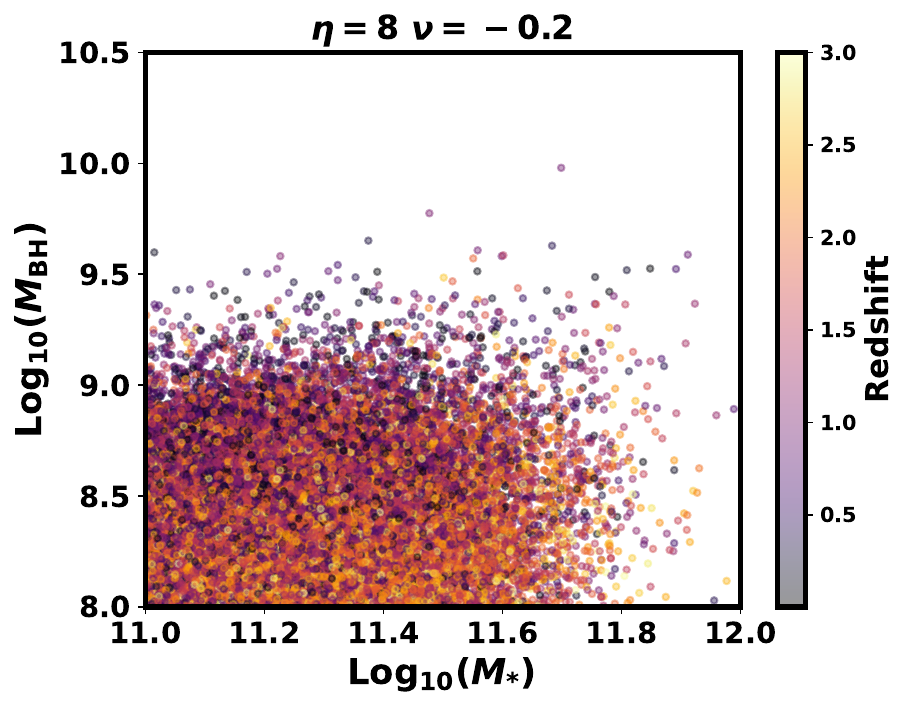}}
    \end{minipage}%
    \begin{minipage}{0.4\linewidth}
        \centering
        \subfigure[]{\label{M1_Ms_4}
        \includegraphics[width=\linewidth]{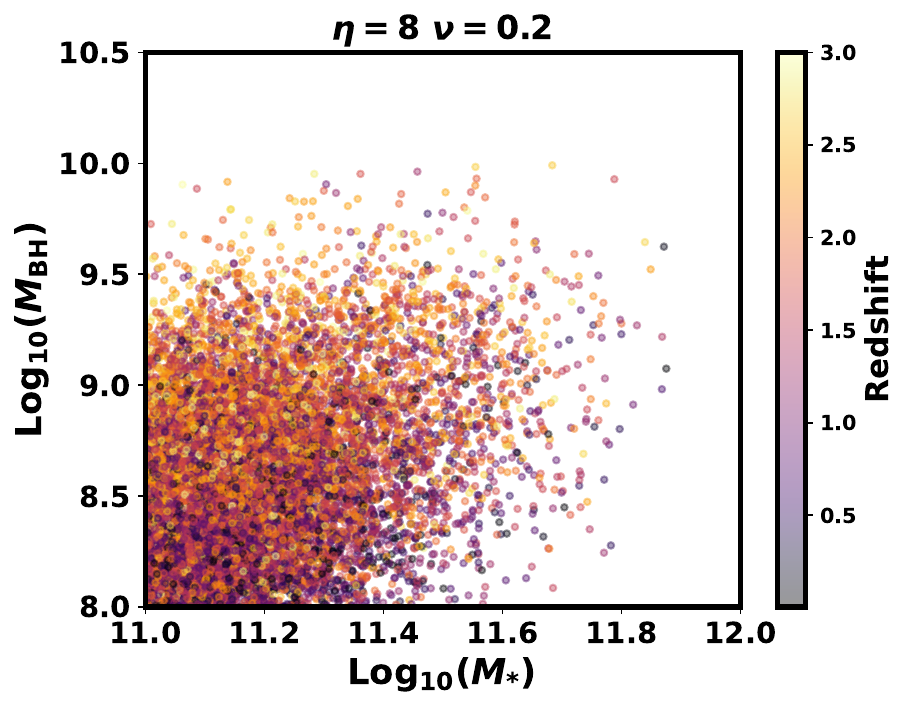}}
    \end{minipage}%
    
    \caption{Plots illustrating the primary masses ($M_{\rm BH}$) of supermassive BHBs as a function of the stellar mass ($M_{*}$) of their host galaxies, derived from simulations. The results are shown for different values of $\eta$ and $\nu$, with redshift represented by the colormap.}
    \label{M1_Ms}    
\end{figure*}

\begin{figure}
    \centering    \includegraphics[width=14cm]{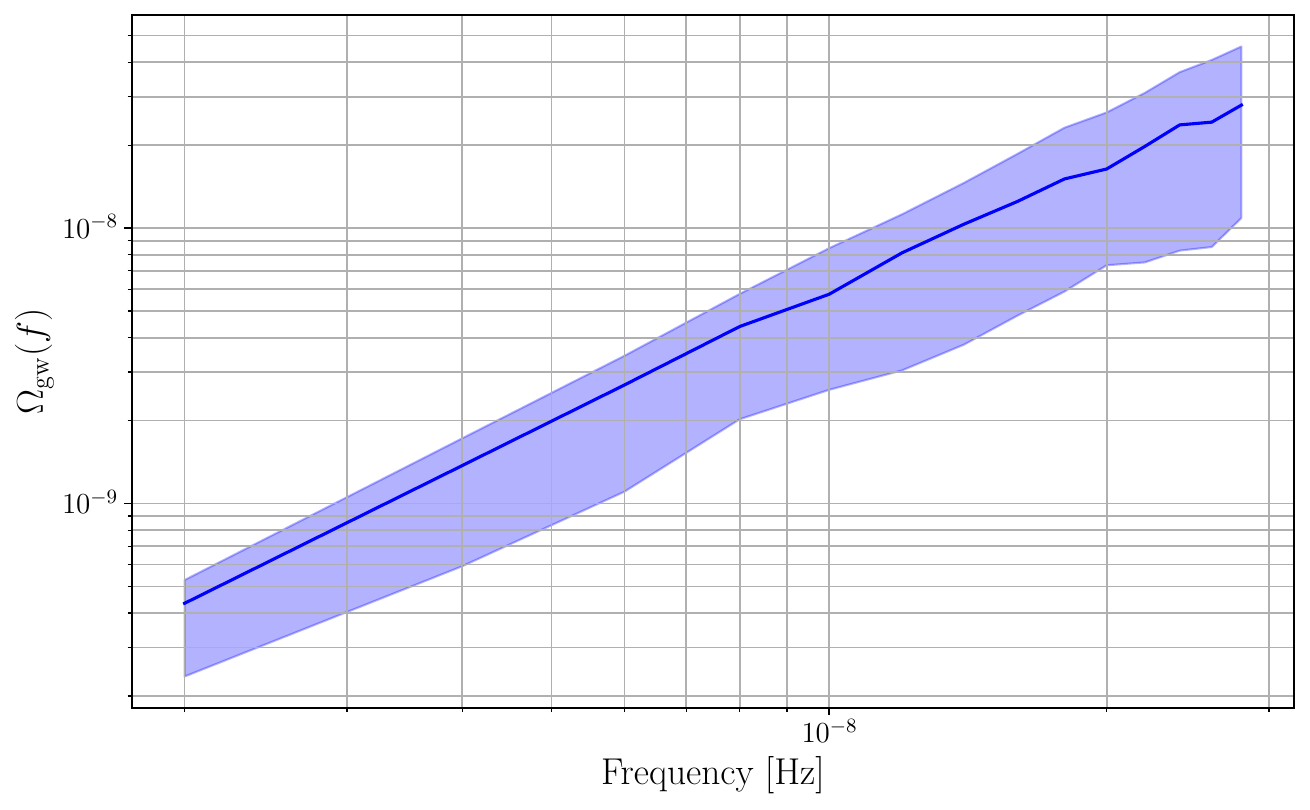}
    \caption{$\Omega_{\rm gw}(f)$ for $\eta = 8.5 ~ \nu = 0$. The solid line represents the median values of the $\Omega_{\rm gw}(f)$ over 1000 realizations, and the shaded region represents the 68\% confidence interval.  }
    \label{fig:Omega}
\end{figure}

\section{Comparison of Multi-tracer Correlated Stacking with the angular power spectra}\label{Comp}
Nearly all studies of anisotropic SGWB rely on the angular power spectrum ($C_{\ell}$) as the primary tool for measuring the signal \citep{taylor2015limits,agazie2023nanograv,sato2024exploring,gardiner2024beyond}. The angular power spectrum ($C_\ell$) measures the variance of the SGWB anisotropies at different angular scales denoted by $\Delta \theta$, which can be approximately related to the multipoles as $\ell\sim 180/\Delta \theta$. It is defined as 
\begin{equation}
C_\ell \equiv \frac{1}{2\ell + 1} \sum_{m=-\ell}^\ell |a_{\ell m}|^2,    
\end{equation}
where $a_{\ell m}$ are the spherical harmonic coefficients of the SGWB fluctuation ($\Delta\Omega_{\rm gw}(f)$). For a shot noise-dominated distribution, which is most likely the case for the nHz SGWB, $C_\ell \propto \Big<|\Delta\Omega_{\rm gw}(f)|^2\Big>$ and exhibit a flat spectrum in the harmonic-space. While $C_{\ell}$ is a valuable estimator, one of the major drawbacks is that it primarily captures the Gaussian component of the signal and may overlook important non-Gaussian features. However, in reality, the SGWB fluctuation ($\Delta\Omega_{\rm gw}(f)$) is expected to be highly non-Gaussian due to a fewer number of sources distributed across the sky \citep{sah2024imprints}. In Fig. \ref{Hist_Cl}, in the top plot we show two distributions: the distribution of angular power spectrum coefficients ($C_{\ell}$) over 1000 realizations of the SGWB for different $\ell$ modes and two frequencies.
The bottom plot illustrates the distribution of spatial SGWB fluctuations ($\Delta\Omega_{\rm gw}$) for 4 specific realizations. As expected for the shot noise-dominated signals, the distribution of $C_{\ell}$ at all $\ell$ modes appears similar. The distribution of $C_{\ell}$ shows a significant spread, reflecting the variability in the anisotropic signal due to the stochastic nature of the supermassive BHB population across different realizations and it also peaks close to the value for isotopic scenario $C_\ell=0$.  Similarly, the distribution of $\Delta\Omega_{\rm gw}$  is highly skewed toward positive values. While $C_{\ell}$ primarily captures the Gaussian components of the anisotropic SGWB, the presence of extended tails in the $\Delta\Omega_{\rm gw}$ distribution indicates a highly non-Gaussian nature. These extended tails suggest that $C_{\ell}$ can misrepresent or underestimate the contributions from non-Gaussian features, potentially leading to a biased or incomplete characterization of the anisotropic signal. Therefore, $C_{\ell}$ is not the best way to quantify the anisotropy of the SGWB.

 In the \texttt{Multi-Tracer Correlated Stacking} method, we directly stack SGWB fluctuations in the regions of galaxy overdensity. By doing so we aim to add the contribution of the region with a positive  SGWB fluctuation. If supermassive BHBs indeed trace the galaxy distribution, the stacking is expected to yield a strongly positive value, reflecting the anisotropic features of the SGWB. Thus, the stacked signal effectively captures the anisotropy by emphasizing regions associated with galaxy overdensity. While $C_\ell$ primarily measures the variance of the signal, the stacking method provides a direct probe of anisotropies associated with the galaxy distribution, making it uniquely sensitive to anisotropic and non-Gaussian distribution, unlike the angular power spectra $C_\ell$. 
 In Fig. \ref{fig:Map1}, we present SGWB maps at two different GW frequencies, and for two different realizations of the GW source population. Additionally, we show the AGN catalog map, which serves as a tracer for the underlying supermassive BHB distribution. High-frequency maps are significantly less populated because the number of sources emitting at higher frequencies is expected to be much smaller than those emitting at lower frequencies. This is due to the shorter residence time of binaries at higher GW frequencies, which reduces the number of sources contributing to the signal in these frequency ranges. Colored square boxes mark regions of high galaxy density in the AGN map and their corresponding regions in the SGWB maps. The maps illustrate significant variations in the SGWB signal both across frequencies and between realizations. These variations highlight how the anisotropic SGWB is influenced by the stochastic nature of the supermassive BHB population. The spatial correlation between high-density regions in the AGN map and enhanced SGWB signals in the corresponding regions of the GW maps highlights the correlation of SGWB with AGN as a tracer and the potential of stacking methods for detecting anisotropic signals.

The effectiveness of this method, however, is inherently dependent on the resolution of the map (maximum value of spherical harmonic mode, $\ell_{\rm max}$). A higher resolution allows for the identification of smaller-scale features that would smooth out in the lower-resolution signals. This enhances the precise identification of regions with high galaxy density and better localization of anisotropic features. In comparison, the angular power spectrum is dominated by shot noise, which leads to a flat power spectrum. As a result, while a higher resolution map improves the signal-to-noise ratio \citep{pol2022forecasting, raidal2024statistics, depta2024pulsar}, it cannot resolve any features in the angular power spectrum that arise from anisotropic distributions of GW sources based on their connection with host galaxies.

\section{Simulation of SGWB from supermassive BHB population}\label{Sim_pop}
In this section, we demonstrate the simulations that can be used to understand the impact of \texttt{Multi-Tracer Correlated Stacking} method. The supermassive BHBs are the most plausible astrophysical source expected to contribute to the nHz SGWB. These binaries, formed during galaxy mergers, emit gravitational waves during their inspiral phase, producing nearly monochromatic signals over the observation period due to their stable orbits. The superposition of GW signals from numerous unresolved supermassive BHBs generates the SGWB signal in the nHz frequency range. 
This background encodes key information about the supermassive BHB population, such as their merger history, mass function, and the environment in which they evolved. Understanding this requires detailed modeling of the supermassive BHB population and its evolution across cosmic time. In the following subsections, we first discuss the theoretical framework for the SGWB from the supermassive BHB population and then describe the simulation methodology used to model the anisotropic signal.

\subsection{SGWB from supermassive BHB population}
Mathematically, the SGWB density is defined as the energy density of GW per unit logarithmic frequency divided by the critical energy density of the Universe \citep{phinney2001practical,sesana2008stochastic,christensen2018stochastic}. It is given by 
\begin{equation}
    \Omega_{\rm gw}(f, \hat{\omega}) = \frac{1}{\rho_c c^2}  \int  \prod_{i}^{n} d\theta_i \int\limits_{z_{ min}}^{\infty} ~dz~ \frac{d^2V}{d\omega dz} 
    \times \bigg[\frac{ d^{n+4}N(z,\Theta_n,\hat{\omega})}{ d\Theta_n dVd\log f_r}\bigg]  \bigg[\frac{1}{4 \pi d_{L}^{2} c} 
    \frac{ dE_{\rm{gw}}(f_r,\Theta_n)}{{ dt_r}} \bigg],
    \label{SGWB2}
\end{equation}
where, $\rho_c$ is the critical density of the Universe, $c$ is the speed of light, $\hat{\omega}$ represents the sky direction. The term $\frac{dN^{n+4}(z, \Theta_n, \hat{\omega})}{d\Theta_n dV d\log f_r}$ represents the comoving number density of the binaries, per unit source parameter space ($\Theta_n$), and per unit logarithmic frequency interval. The factor $\frac{dE_{\rm gw}(f_r, \Theta_n)}{dt_r}$ is the luminosity of the source. $\frac{dN^{n+4}(z, \Theta_n, \hat{\omega})}{d\Theta_n dV d\log f_r}$, can be modeled based on the relationship between supermassive BHBs and their host galaxy properties. It can be expressed as

\begin{equation}
    \frac{d^7N}{dM_{\rm BH} dq dV \log f_r} \propto \int dM_{*} \frac{d^4N}{dM_* dV} P(M_{\rm BH} | M_*, z) P(q | M_*, z) P(f_r | M_*, z),
    \label{eq:binary_population}
\end{equation}
where $P(M_{\rm BH} | M_*, z)$ is the probability distribution of primary BH masses conditioned on the galaxy stellar mass ($M_{*}$) and redshift, $P(q | M_*, z)$ represents the distribution of mass ratios ($q$) of the binary, given $M_*$ and $z$, and $P(f_r | M_*, z)$ describes the distribution of GW frequencies emitted by binaries, which is directly related to their orbital separation. For simplicity, we adopt parametric forms for $P(M_{\rm BH} | M_*, z)$ and $P(q | M_*, z)$ as

\begin{equation}
    P( M_{\rm BH}|  M_{*},z) \propto \mathcal{N}\Big(\mathrm{Log}_{10}\Big(\frac{M_{\rm BH}}{M_{\odot}}\Big)| \mathrm{Log}_{10}\Big(\frac{M_{\mu}(M_*, z)}{M_{\odot}}\Big),\sigma_m \Big),
    \label{M1}
\end{equation}

\begin{equation}
     P(q|  M_{*},z) \propto \bigg\{
    \begin{array}{cl}
    & 1/q, \quad  0.01 < q < 1,\\
    & 0, ~~ \rm else ,
    \end{array}
    \label{q}
\end{equation}
where $\mathcal{N}$ denotes a normal distribution with mean $M_{\mu}(M_*, z)$ and standard deviation $\sigma_m$. The mean BH mass is modeled as \citep{sah2024imprints}

\begin{equation}    \mathrm{Log}_{10}\Big(\frac{M_{\mu}(M_*, z)}{M_{\odot}}\Big) = \eta + \rho~ \log(M_*/10^{11} M_\odot) + \nu~ z,
    \label{Mmu}
\end{equation}
where parameters $\eta$, $\rho$, and $\nu$ define the scaling relationship between BH mass, and host galaxy stellar mass, as a function of redshift. The parameter $\eta$ sets the normalization, representing the typical supermassive BH mass for a galaxy with stellar mass $M_* = 10^{11} M_\odot$ at redshift $z = 0$. The parameter $\rho$ encodes the correlation between BH mass and stellar mass. Finally, $\nu$ governs the redshift evolution of this relation.

\subsection{Simulation Methodology}

In this work, we compute the SGWB by performing Monte Carlo simulations of the supermassive BHB population, building upon the method outlined in \cite{sah2024imprints}. In this technique, supermassive BHBs contributing to the PTA signal are assigned to the galaxies in the galaxy catalog by considering the stellar mass–supermassive BH mass ($M_{*} - M_{\rm BH}$) relations described in Eqs.~\eqref{M1} to \eqref{Mmu}. We only assign binaries to galaxies where the mean primary BH mass, as given by Eq. \eqref{Mmu}, exceeds $10^{8} M_{\odot}$. The occupation fraction of galaxies is determined by comparing the SGWB generated by the binary population to the assumed SGWB power spectrum. Sources are sampled from the catalog such that the total SGWB generated by all binaries in each frequency bin aligns with the expected SGWB power spectrum. 

In this work, we employ an AGN catalog, \texttt{THE MILLION QUASARS (MILLIQUAS) CATALOGUE}, introduced in \cite{flesch2023million}. The AGN catalog offers a significant advantage due to its completeness at higher redshifts, which allows us to more accurately account for the supermassive BH population contributing to the SGWB over larger redshifts. However, a key limitation of the catalog is its lack of stellar mass information. To address this, we employ the redshift-dependent Schechter function, following \cite{mcleod2021evolution}, to assign stellar masses to the AGNs. Stellar masses are assigned through Monte Carlo sampling from the Schechter function, under the assumption that these AGNs are potential hosts of supermassive BHBs. Specifically, we consider only AGNs with stellar masses exceeding the minimum threshold required to host supermassive BHBs of primary mass $>10^{8} M_{\odot}$ as determined by Eq. \eqref{Mmu}. {In the simulations, we assume that a fraction of the SGWB power spectrum (denoted by Frac) is contributed by AGNs. This fraction of $\Omega_{\rm gw}(f)$ is generated through Monte Carlo sampling of sources from the AGN catalog. The remaining portion (1-Frac) of the SGWB is generated by randomly and uniformly distributing sources across the sky, ensuring they follow the same redshift distribution as the catalog and adopt the Schechter function for stellar mass. 

Since AGNs are strongly associated with galaxy mergers and are probably more likely to host SMBHBs \citep{de2019quest,casey2022quasar,saeedzadeh2024dual}, this technique can be extended to investigate the fraction of SMBHBs residing in AGNs. Such an analysis offers a novel approach to studying the formation and evolution of both supermassive BHBs and AGNs. By integrating stacking results with other multimessenger studies, such as the nHz SGWB and periodic AGN candidate analyses \citep{casey2024quasars}, we can further refine our understanding of the astrophysical sources contributing to the SGWB.}

In Fig.~\ref{M1_Ms}, we present the primary supermassive BH mass ($M_{\rm BH}$) as a function of the host galaxy stellar mass ($M_{*}$) for different values of $\eta$ and $\nu$, based on the simulation described above. The color bar represents the redshift of the source. The mass of the primary supermassive BH is restricted to $M_{\rm BH} \geq 10^{8} M_{\odot}$, as lower-mass BHs are unlikely to contribute significantly to the GW background.
In Figs. \ref{M1_Ms_1}, and \ref{M1_Ms_2}, where $\nu = 0$, the supermassive BH mass-stellar mass ($M_{\rm BH} - M_{*}$) relation is independent of redshift. In contrast, Figs. \ref{M1_Ms_3} and \ref{M1_Ms_4} illustrate cases with $\nu = -0.2$ and $\nu = 0.2$, respectively, indicating an redshift evolving $M_{\rm BH}$-$M_{*}$ relation. A negative $\nu$ ($\nu = -0.2$) implies less massive supermassive BHs at higher redshifts, as shown in Fig. \ref{M1_Ms_3}, while a positive $\nu$ ($\nu = 0.2$) suggests more massive supermassive BHs at higher redshifts occupying similar mass galaxies, as observed in Fig. \ref{M1_Ms_4}. This evolution is clearly reflected in the plots: for $\nu = -0.2$, the high-redshift supermassive BHs lie below the low-redshift supermassive BHs for a given $M_{*}$, whereas the trend reverses for $\nu = 0.2$, with high-redshift supermassive BHs lying above their low-redshift counterparts for the same $M_{*}$.
In Fig. \ref{fig:Omega}, we show the median value of the $\Omega_{\rm gw}(f)$ along with its fluctuation over 1000 realizations. These fluctuations arise from the normalization process of the SGWB in each realization. Specifically, supermassive BHBs are randomly sampled from the catalog and added iteratively until the SGWB density in each frequency bin is equal to or slightly below the presumed value. As a result, the SGWB density in each frequency bin is either equal to or slightly smaller than the presumed value.

\section{Multi-Tracer Correlated Stacking Formalism}\label{Form}

The \texttt{Multi-Tracer Correlated Stacking} technique is designed to detect anisotropies in the SGWB by leveraging the correlation between the distribution of tracers of supermassive BHBs and the SGWB map, stacking contributions from regions of the sky with higher densities of these tracers. The central idea is that supermassive BHBs, which are believed to be the dominant contributors to the nHz SGWB, reside within massive galaxies. As a result, the SGWB signal should trace the distribution of galaxies, particularly in regions of high galaxy density. The anisotropy of the SGWB can be demonstrated by establishing that the SGWB signal is predominantly sourced from these high-density regions. 

We define the relative fluctuation of galaxy density at a sky position $\hat{\omega}$ as
\begin{equation}
\delta_g(\hat{\omega}) = \frac{n_g(\hat{\omega}) - \bar{n}_g}{\bar{n}_g}, 
\end{equation}
and relative fluctuation in the SGWB energy density, \( \Omega_{\rm gw}(f, \hat{\omega}) \), can be defined as
\begin{equation}
    \Delta\Omega_{\rm gw}(f,\hat{\omega}) \equiv  \frac{\Omega_{\rm gw}(f,\hat{\omega}) - \overline{\Omega}_{\rm gw}(f)}{\overline{\Omega}_{\rm gw}(f)}, 
    \label{SGWB_Fluc}
\end{equation}
where \(n_g(\hat{\omega})\) is the number density of galaxies in the direction $\hat{\omega}$. $\bar{n}_g$ and $\overline{\Omega}_{\rm gw}(f)$ are the mean number density of galaxies and mean SGWB density, respectively, across all pixels. A positive fluctuation \(\delta_g(i) > 0\) or $\Delta\Omega_{\rm gw}(f,\hat{\omega})>0$ indicates an overdensity of galaxies or SGWB respectively in that pixel. The uncertainty in the measurement of the $\Delta\Omega_{\rm gw}$ is given by the diagonal elements of the inverse of the Fisher information matrix, $\rm R^{T} \Sigma^{-1} R$ (see Appendix. \ref{sec:appendixA}), where R is the overlap response matrix of the pulsar pairs, and $\Sigma$ is the covariance matrix of the amplitude scaled optimal estimator ($\hat{\textbf{S}}_{IJ}$) of cross-correlation between the timing of pair of pulsars I and J. $\Sigma$ is the diagonal matrix with diagonal elements given by $\bar{\sigma}_{IJ}^{2}$, representing the uncertainty in the measurement of the timing cross-correlation between the pulsars I and J.

For the \texttt{Multi-Tracer Correlated Stacking}, we select pixels where the relative galaxy density fluctuation exceeds a certain threshold $\delta_g(i) > \delta_{\text{cut}}$, where $\delta_{\text{cut}}$ is a chosen cutoff value. 

We set $ \delta_{\rm cut} = 0 $, which corresponds to the scenario where SMBHBs reside in regions of galaxy overdensity. Cases with $ \delta_{\rm cut} > 0 $ are particularly relevant when SMBHBs are expected to be predominantly located in regions with very high galaxy number density, allowing for a more targeted selection of environments that are most likely to host these systems. Once the pixels with overdensities are identified, the corresponding pixels from the SGWB map are stacked. This stacking process involves summing the SGWB fluctuation signals from the selected pixels, effectively increasing the contribution from regions with high galaxy density. The stacked SGWB signal can be expressed as

\begin{equation}
    \hat{\Omega}_{\text{stacked}} \equiv  \sum_{i \in  \delta^{i}_g > \delta_{\text{cut}}} \Delta\Omega_{\text{gw}}^{i}
    \label{stack},
\end{equation}
where index i represents the pixel index.

This technique not only helps in detecting anisotropies but also provides a direct way to test whether the SGWB traces the large-scale distribution of AGNs, quasars, and galaxies. 
Unlike the cross-correlation ($C_{\ell}^{\rm gGW}$) between the galaxy and SGWB, which measures statistical correlations between two fields (e.g., galaxy distribution and SGWB map) averaged over all angular scales, the stacking technique takes a more targeted approach. By summing the contributions from specific regions where the supermassive BHB population is expected to dominate, the stacking method amplifies the anisotropic signal, focusing directly on the alignment between the galaxy density and the SGWB.

We demonstrate in the next section that by applying this formalism, we can enhance the sensitivity of pulsar timing arrays (PTAs) to discover SGWB anisotropies. This approach offers a novel and complementary method to other traditional methods like angular power spectrum estimators. The optimal estimator of SGWB density amplitude ($\Omega_{\rm gw}(f_{\rm ref})$) and the corresponding uncertainty in its measurement ($\sigma_{IJ}$) using timing residual of pairs of pulsars (I,J) are derived in the Appendix. \ref{sec:appendixA}.

\section{Demonstration of the Method on the Simulated population of supermassive BHB using AGN catalog}\label{App}

In this section, we apply the \texttt{Multi-tracer Correlated Stacking} technique to a simulated population of supermassive BHBs obtained using the method described in Sec. \ref{Sim_pop}, based on an AGN catalog. The AGN catalog, unlike other galaxy catalogs, provides a complete sample of active galaxies across a wide range of redshifts. This enables more accurate modeling of the distribution of supermassive BHBs contributing to the SGWB. For the stacking process, we select pixels in the sky where the galaxy density fluctuation, $\delta_{\rm g}(i) > 0$ (i.e., $\delta_{\rm cut} = 0$). The SGWB fluctuation ($\Delta\Omega_{\rm gw}$) at these pixels is then summed to produce the stacked SGWB signal ($ \hat{\Omega}_{\text{stacked}}$).

We consider different numbers of pulsars ($N_{\rm p}$) and sky resolutions for the analysis. For fiducial cases, the pulsars are assumed to be isotropically distributed, while we also present a case using the realistic distribution of NANOGrav pulsars to demonstrate its impact on the results. For this analysis, we take the $ \bar{\sigma}_{IJ} = 1$ which is close to the median uncertainty in the measurement of the $\hat{S}_{IJ}$ for current PTA with $T_{\rm obs} = 20$ years. To model the amplitude of the isotropic SGWB signal, we fit a simple power-law model, $A \times (f/f_{\rm yr})^{\alpha}$, to $\Omega_{\rm gw}(f)$ estimate of the NANOGrav 15-year dataset \citep{agazie2023nanograv} using the PTArcade code \citep{mitridate2023ptarcade}. The parameters $A$  and $\alpha$ are determined from the fit, and we assume the median values of these fitted parameters to compute the isotropic SGWB energy density $\Omega^{\rm iso}_{\rm gw}$, which is used as a reference for the rest of the analysis.

\begin{figure}
  \centering
  \subfigure[]{\label{fig:stack_Iso0.5}
    \includegraphics[width=0.8\linewidth,trim={0.cm 0  0 0.cm},clip]{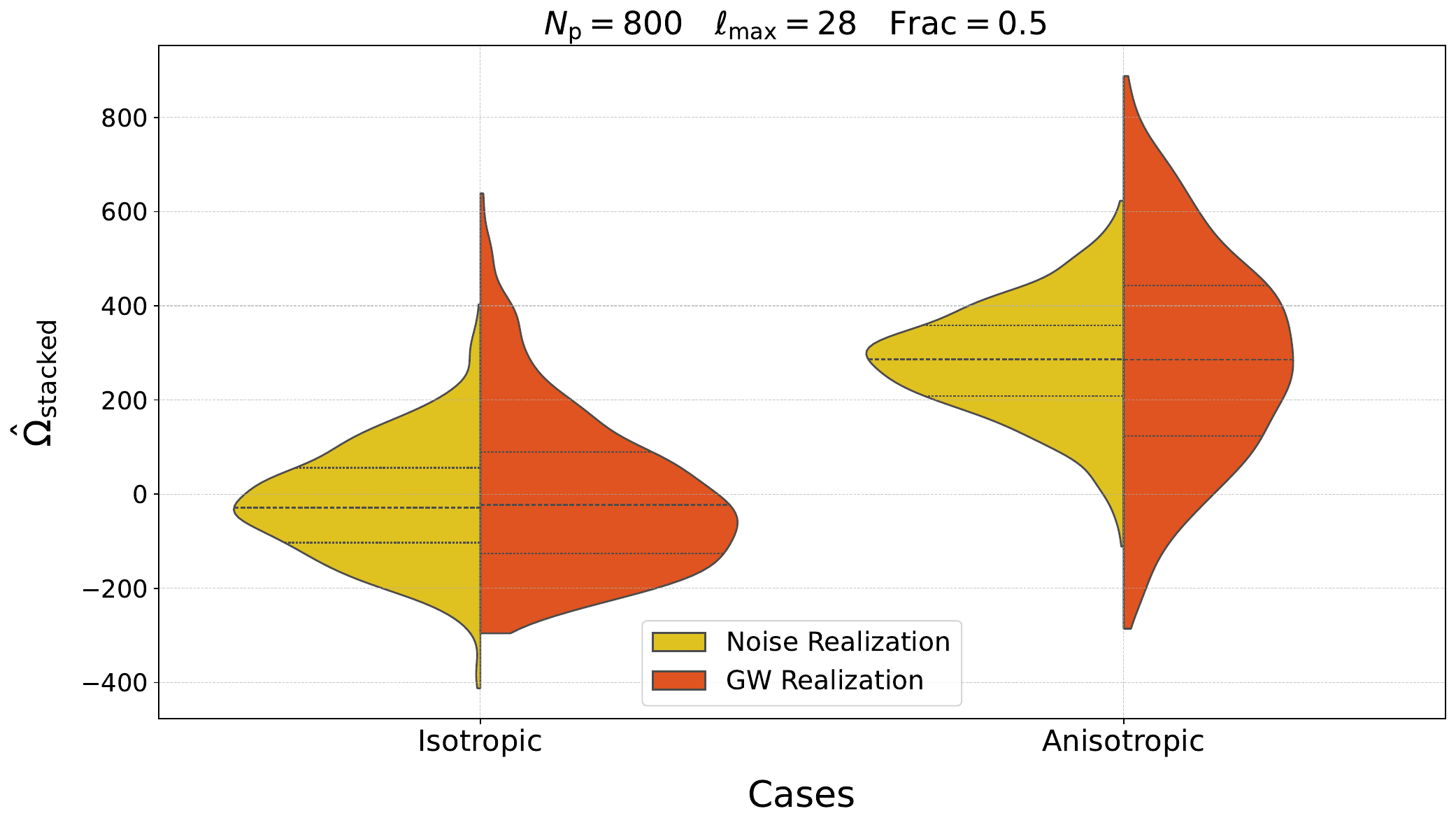}}
  \vspace{0.5cm}
  \subfigure[]{\label{fig:stack_Iso1}
    \includegraphics[width=0.8\linewidth,trim={0.cm 0  0 0.cm},clip]{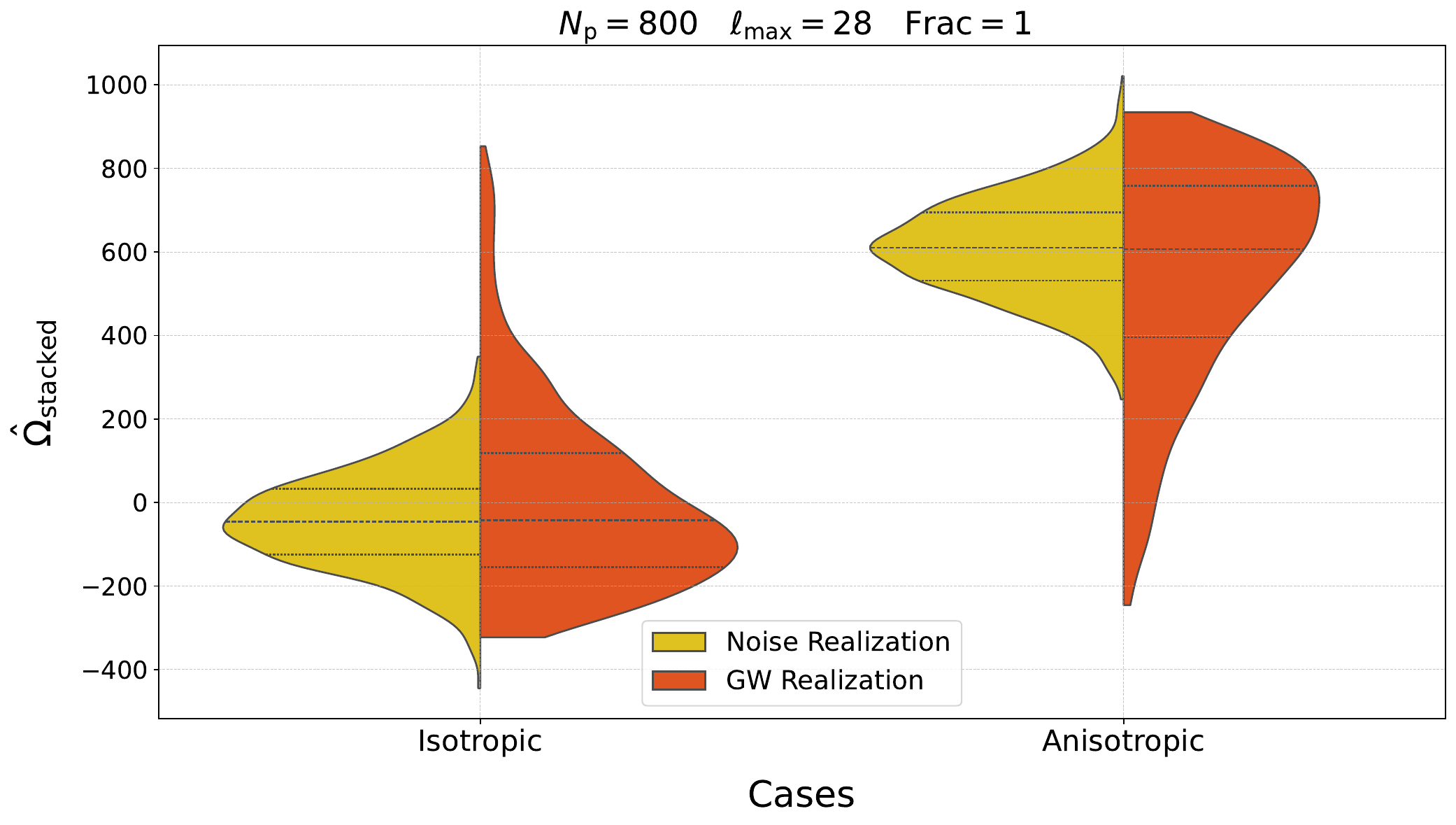}}
  \caption{Violin plots of the stacked SGWB signal ($\hat{\Omega}_{\text{stacked}}$) for $\eta = 8.5$ and $\nu = 0$, comparing two scenarios: supermassive BHBs residing at the centers of galaxies/AGNs (anisotropic case) and isotropically distributed supermassive BHBs (isotropic case). The results are shown for (a) an AGN fraction of 0.5 and (b) an AGN fraction of 1, representing their respective contributions to the overall SGWB density.}
  \label{fig:stack_Iso}
\end{figure}

In Fig. \ref{fig:stack_Iso}, we present a violin plot of the stacked SGWB signal ($\hat{\Omega}_{\text{stacked}}$) for 800 pulsars and $\ell_{\rm max} = 28$, assuming $\eta = 8.5$, $\rho = 1$, and $\nu = 0$.
The figure compares the stacked signal for two distinct scenarios: some fraction of supermassive BHBs contributing to the SGWB if hosted in AGN and isotropically distributed supermassive BHBs. The AGN contribution to the SGWB density is parameterized as $\Omega^{\rm AGN}_{\rm gw} = \text{Frac} \times \Omega^{\rm iso}_{\rm gw}$. Figs. \ref{fig:stack_Iso0.5} and \ref{fig:stack_Iso1} correspond to scenarios with $\text{Frac} = 0.5$ and $\text{Frac} = 1$, respectively. The plot shows the uncertainty in the signal due to the noise as well as the variation of the signal across 1000 realizations of the GW population. The noise realization is centered around the median value of the stacked signal obtained from the 1000 GW population realizations. This configuration serves as the fiducial setup for the rest of the analysis.

The stacked signal for the AGN-centered supermassive BHB population exhibits significantly positive values, reflecting the correlation between the GW source and AGN distribution. In contrast, the isotropically distributed supermassive BHB population shows a stacked signal centered around zero, indicative of an isotropically distributed source without correlation with AGN/galaxy distribution. This result underscores the capability of the \texttt{Multi-tracer Correlated Stacking} method to distinguish between an isotropic distribution and an anisotropic distribution of GW sources that follow the galaxy distribution in the Universe. Unlike traditional methods of angular power spectra, this technique is uniquely sensitive to non-Gaussian anisotropic signals.
In the following, we describe how different physical scenarios can influence the strength of the stacked signal.

\begin{enumerate}
    \item \textit{\textbf{Impact of the AGN fraction :}} In Fig. \ref{fig:stack_frac}, we illustrate the violin plot of the stacked SGWB signal for four different values of the parameter Frac: 0.25, 0.50, 0.75, and 1. Each case represents the stacked signal under a different assumption about the overall contribution of AGN to the SGWB. Since only a Frac of $\Omega^{\rm iso}_{\rm gw}(f)$ originates from AGNs, only this fraction of the SGWB will follow the AGN distribution.
For smaller values of Frac, a significant portion of high-density pixels in the SGWB map is likely to align with underdense regions in the AGN map. This mismatch reduces the stacked signal, leading to a lower $\hat{\Omega}_{\rm stacked}$. Conversely, for larger values of Frac, the SGWB better traces the AGN distribution, and fewer high-density SGWB pixels align with AGN underdense regions. As a result, the stacked signal becomes more prominent, with a smaller fraction of the distribution falling below zero. This trend is evident in the plot, where the $\hat{\Omega}_{\rm stacked}$ signal for larger Frac values exhibits more positive values. In particular, for Frac = 1, the signal is predominantly positive, indicating a strong alignment between the AGN distribution and the SGWB. This highlights the importance of the AGN fraction in shaping the stacked signal and demonstrates how the stacking technique is sensitive to the degree of correlation between the AGN distribution and the SGWB anisotropy. These results emphasize the utility of the stacking method in probing the relative contribution of AGNs to the SGWB and its connection to astrophysical source distributions.

\begin{figure}
    \centering    \includegraphics[width=15cm]{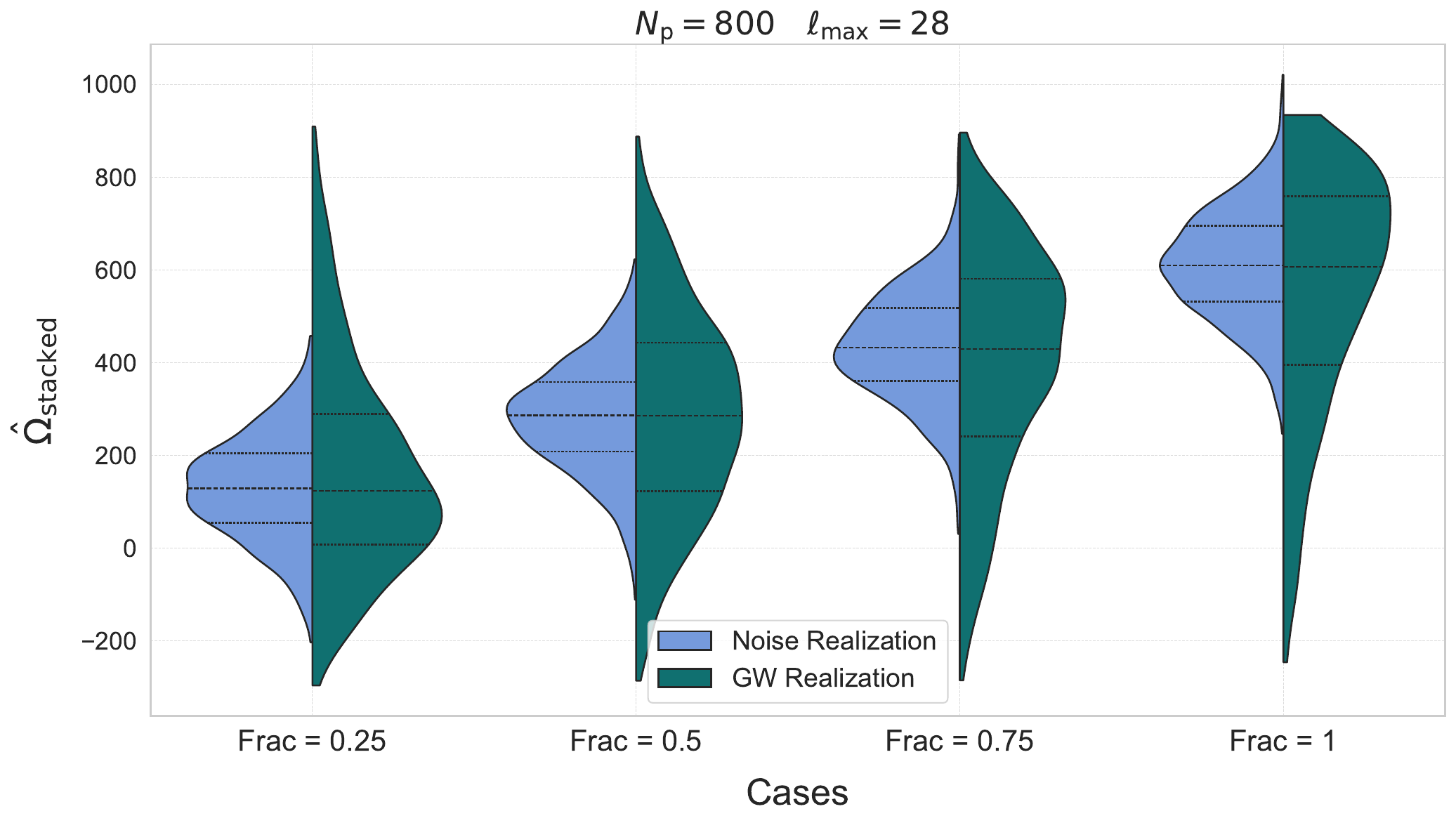}
    \caption{Violin plots of the stacked SGWB signal ($\hat{\Omega}_{\text{stacked}}$) for different contributions of the AGNs to the overall $\Omega_{\rm gw}^{\rm iso}(f)$ signal. The left distribution represents the variance due to noise, while the right distribution shows the variation arising from different GW realizations.
}
    \label{fig:stack_frac}
\end{figure}

\item\textit{\textbf{Impact of the supermassive BHB population model:}} In Fig. \ref{fig:stack_astro}, we present a violin plot of the $\hat{\Omega}_{\rm stacked}$ for various astrophysical scenarios, illustrating the uncertainty in the measurement of the signal caused by noise and the fluctuations in the GW population across 1000 realizations. The result is presented for two different $\Omega_{\rm gw}^{\rm AGN}(f)$ corresponding to Frac = 0.5 and Frac = 1. The left and right distributions represent the uncertainty on the stacked signal caused by noise, and GW source population realizations respectively. The broader spread of the right posterior indicates the significant impact of fluctuation due to the GW realization on the stacked signal. The distribution of all the models is very similar within the fluctuation in the distribution due to different GW realizations.

\begin{figure}
  \centering
  \subfigure[]{\label{fig:stack_astro0.5}
    \includegraphics[width=0.8\linewidth,trim={0.cm 0  0 0.cm},clip]{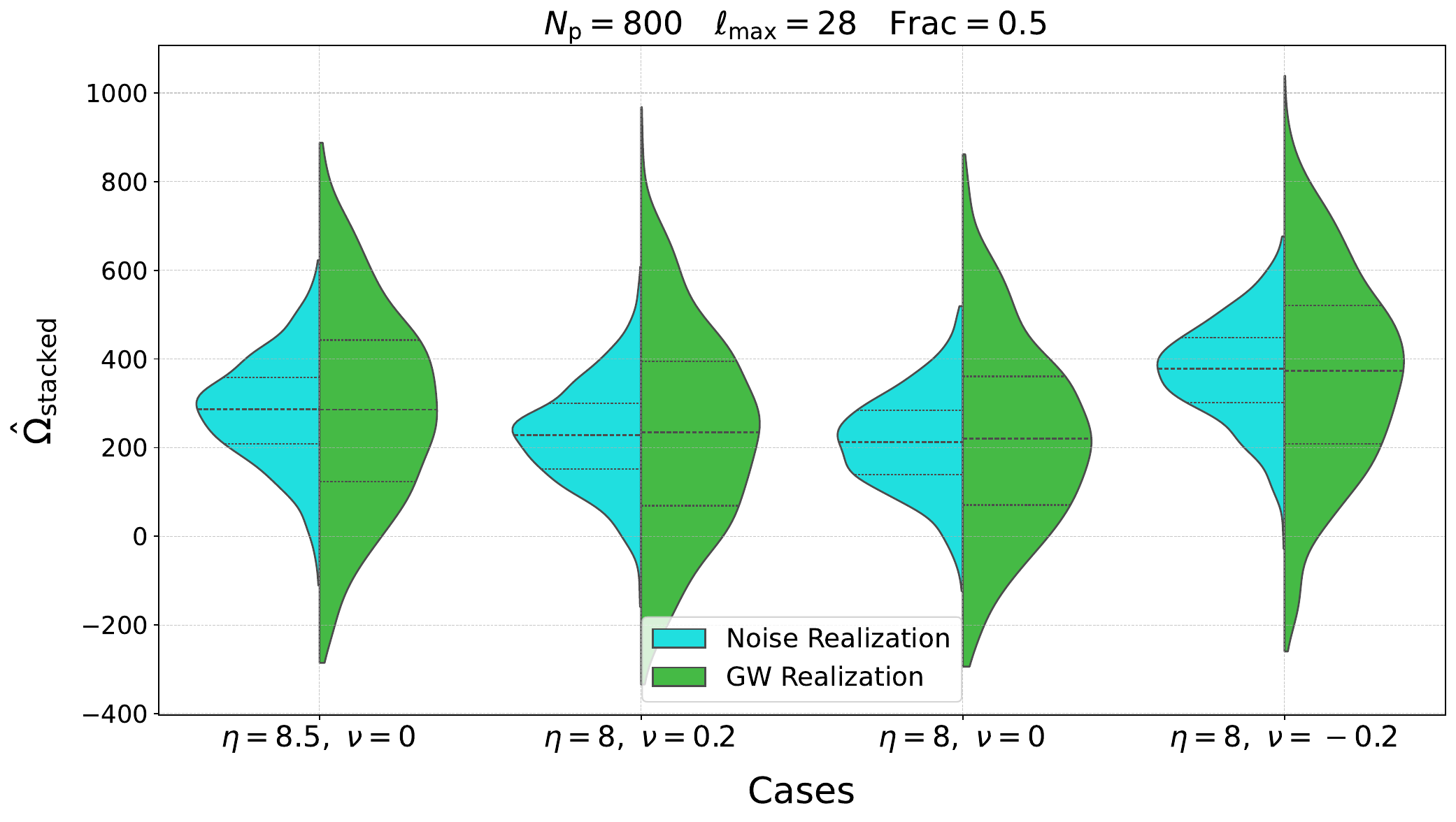}}
  \vspace{0.5cm}
  \subfigure[]{\label{fig:stack_astro1}
    \includegraphics[width=0.8\linewidth,trim={0.cm 0  0 0.cm},clip]{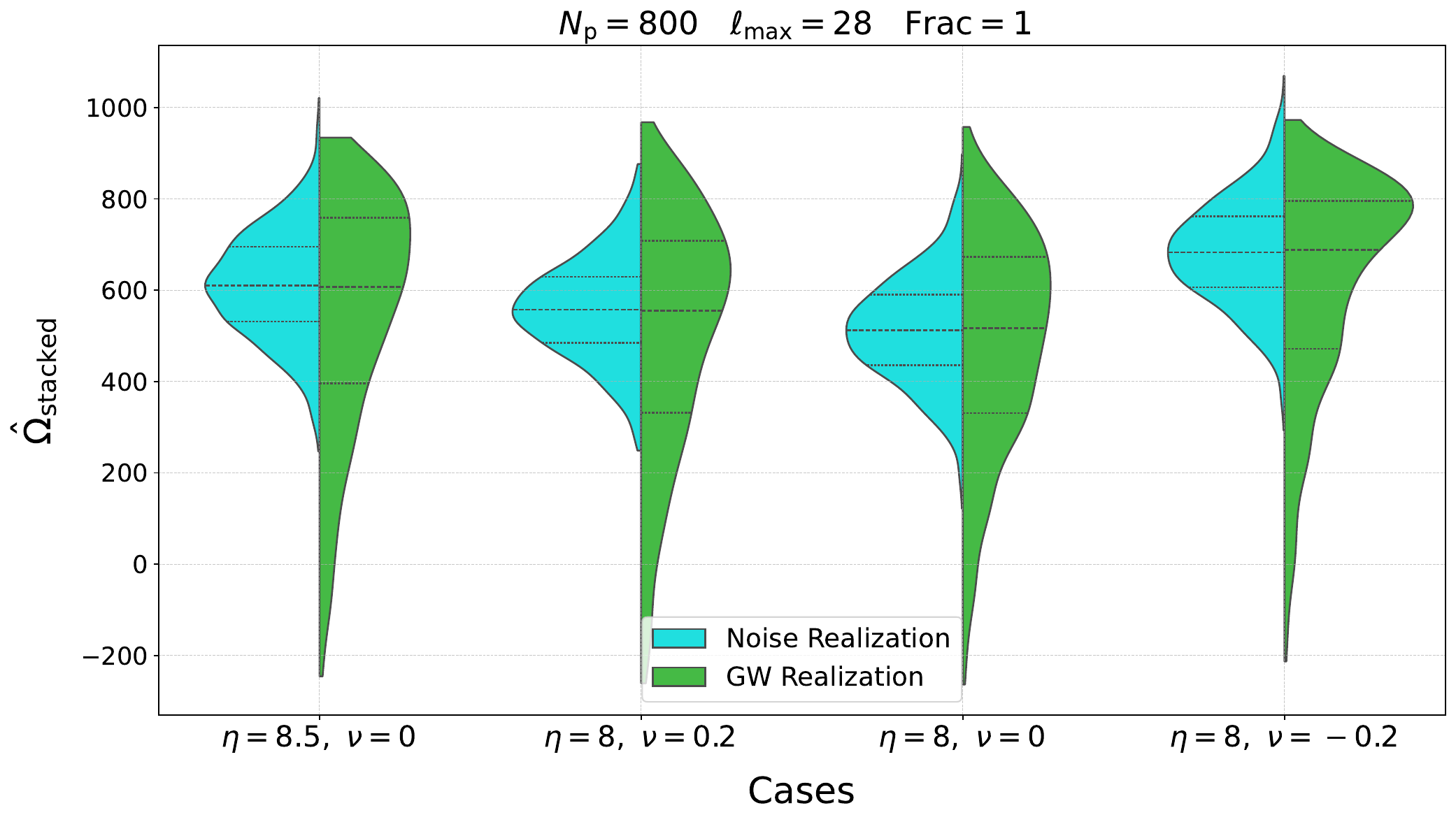}}
  \caption{Violin plots of the stacked SGWB signal ($\hat{\Omega}_{\text{stacked}}$) for different astrophysical scenarios (different values of $\eta$ and $\nu$ ), assuming (a) SGWB with an AGN fraction of 0.5, and (b) SGWB with an AGN fraction of 1. The left distribution represents the variance due to noise, while the right distribution shows the variation arising from different GW realizations.}
  \label{fig:stack_astro}
\end{figure}

\item\textit{\textbf{Impact of the number of pulsars and angular resolution of SGWB map:}} In Fig. \ref{fig:stack_pul}, we show the similar violin plot for the distribution stacked signal for different numbers of isotropically distributed pulsars ($N_{\rm p}$) with corresponding $\ell_{\rm max}$  determined as $\ell_{\rm max}$ $\sim$ $\sqrt{N_{\rm p}}$  \citep{boyle2012pulsar,romano2017detection}. Fig. \ref{fig:stack_pul0.5} and Fig. \ref{fig:stack_pul1} represent the cases with Frac = 0.5 and Frac = 1, respectively. This plot highlights the dependence of the stacked signal on the number of pulsars and the angular resolution. The stacked signal increases as the resolution is improved. This is because higher resolution allows for more precise identification and stacking of contributions from regions with significant anisotropic features in the SGWB signal. Improving the resolution enables us to divide the sky into smaller angular regions. This higher resolution enables the detection of smaller-scale anisotropies that would otherwise be smoothed out in lower-resolution analyses. As a result, the stacked signal better reflects the correlation of sources with the underlying tracer.

To gauge the effectiveness of the \texttt{Multi-Tracer Correlated Stacking}, it is important to understand how likely it is to get a stacked signal less than or equal to zero, even when the SGWB signal is anisotropic and is traced by AGNs. For this, we calculate the p-value for a given anisotropic distribution.

In Fig. \ref{fig:p-value}, we show the p-values of the $\hat \Omega_{\rm stacked}$ distribution for $N_{\rm p} = 600$ and $\ell_{\rm max} = 24$ for two different values of Frac. Here, the p-value is defined as the fraction of the stacked signal less than or equal to zero. We compute the p-values separately for two scenarios: noise realizations with a fixed median value of the $\hat \Omega_{\rm stacked}$ across GW population realizations, and GW population realizations without noise. In Table \ref{tab:table1}, we provide the p-values for various $N_{\rm p}$ and their corresponding $\ell_{\rm max}$, for Frac = 0.5 and Frac = 1. For noise realizations, we observe that the p-value becomes negligibly small beyond a resolution of $\ell_{\rm max} = 20$ for both values of Frac. For GW population realizations, the p-value is 0.140 (0.056) for Frac = 0.5 (Frac = 1) at $\ell_{\rm max} = 8$ and decreases gradually with increasing resolution. This suggests that the likelihood of measuring a negative value of $\hat \Omega_{\rm stacked}$, given an anisotropic background, becomes negligible beyond a certain resolution when ignoring the uncertainty from GW population realizations. Although a non-zero probability of negative values remains even at high resolutions due to GW population variations, this likelihood is small and does not significantly affect the effectiveness of the stacking method for practical use.

\begin{figure}
  \centering
  \subfigure[]{\label{fig:stack_pul0.5}
    \includegraphics[width=0.8\linewidth,trim={0.cm 0  0 0.cm},clip]{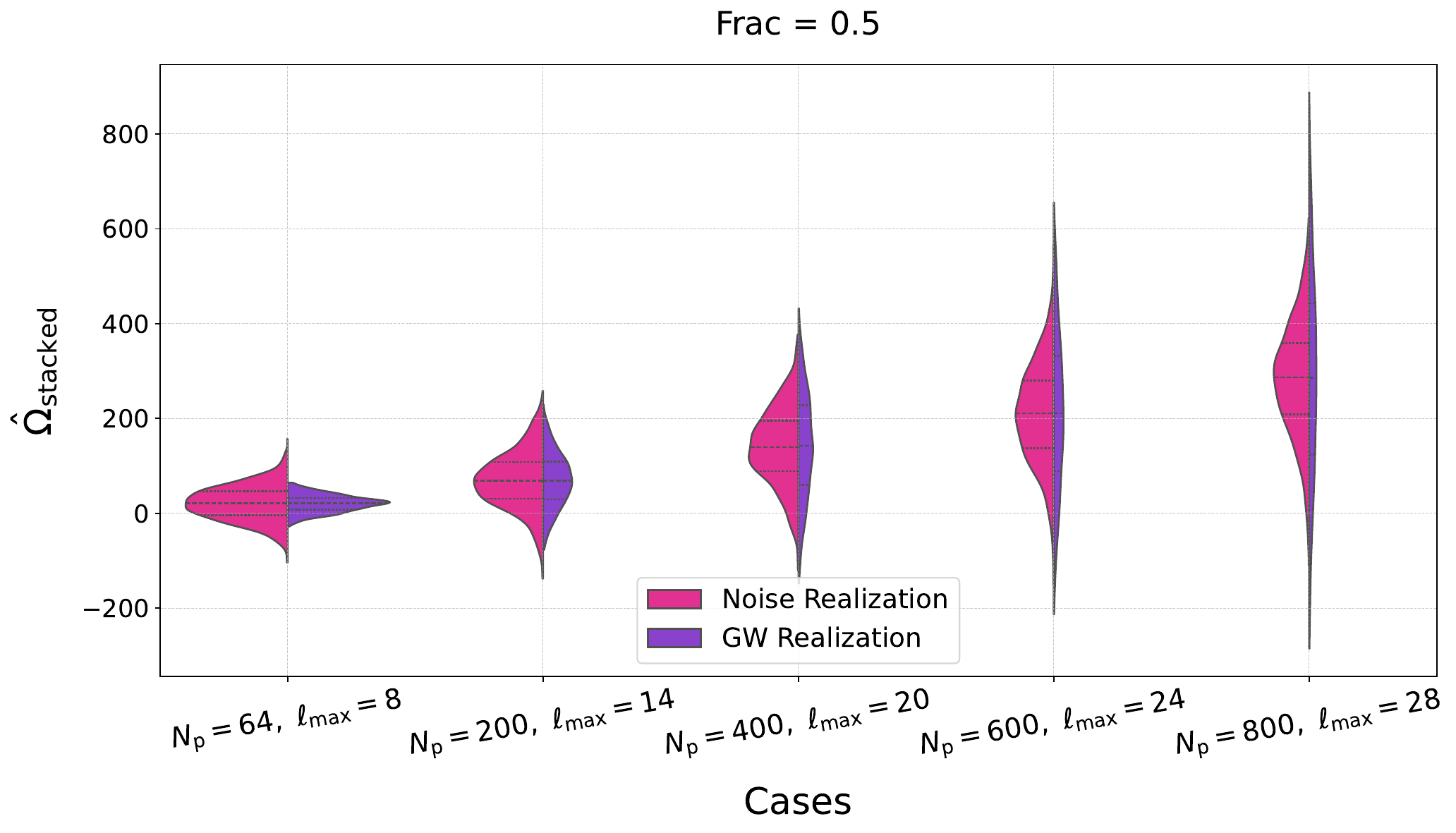}}
  \vspace{0.5cm}
  \subfigure[]{\label{fig:stack_pul1}
    \includegraphics[width=0.8\linewidth,trim={0.cm 0  0 0.cm},clip]{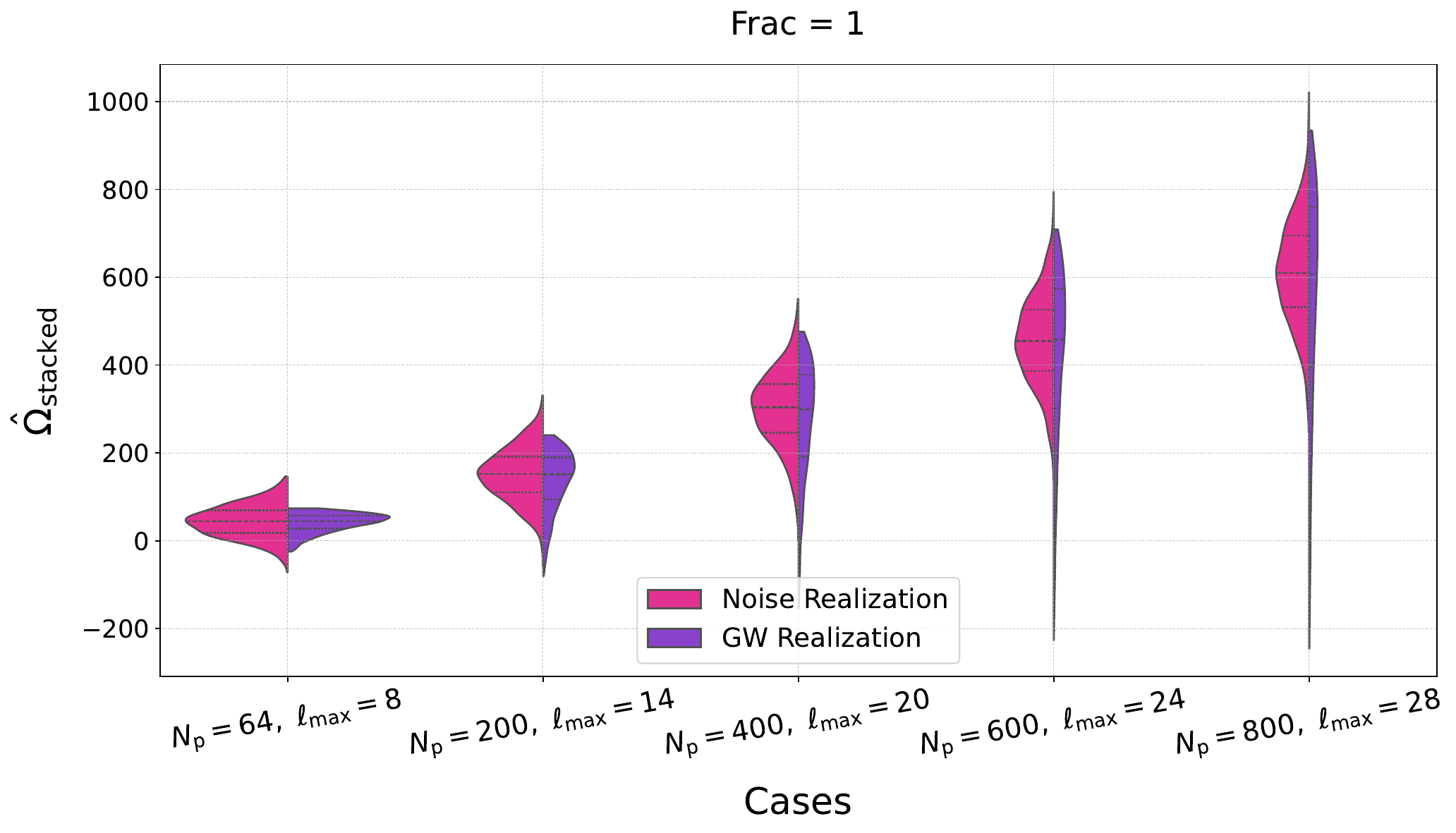}}
  \caption{Violin plots of the stacked SGWB signal ($\hat{\Omega}_{\text{stacked}}$) for different numbers of pulsars ($N_{\rm p}$) and corresponding maximum multipole moments ($\ell_{\rm max}$), assuming (a) SGWB with an AGN fraction of 0.5, and (b) SGWB with an AGN fraction of 1. The left distribution illustrates variations induced by noise, while the right distribution captures the impact of different GW source realizations.}
  \label{fig:stack_pul}
\end{figure}

\begin{figure}
    \centering    
    \includegraphics[width=14cm]{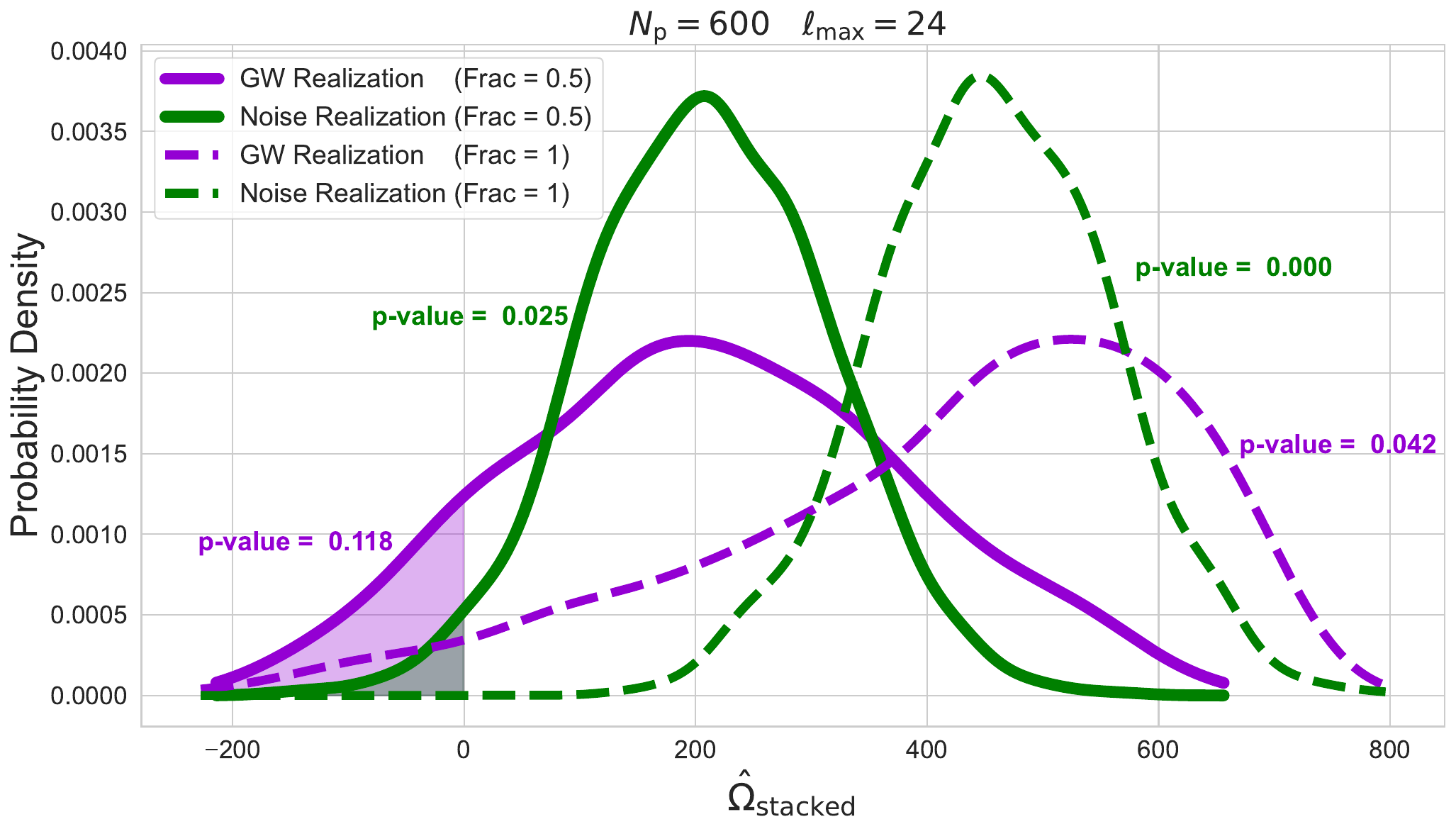}
    \caption{p-values of the $\hat{\Omega}_{\text{stacked}}$ distribution for $N_{\rm p} = 400$ and $\ell_{\rm max} = 24$, assuming an SGWB with AGN fractions of 0.5 and 1. The p-value is defined here as the fraction of the signal less than or equal to zero.}
    \label{fig:p-value}
\end{figure}

\begin{figure}
    \centering    \includegraphics[width=15cm]{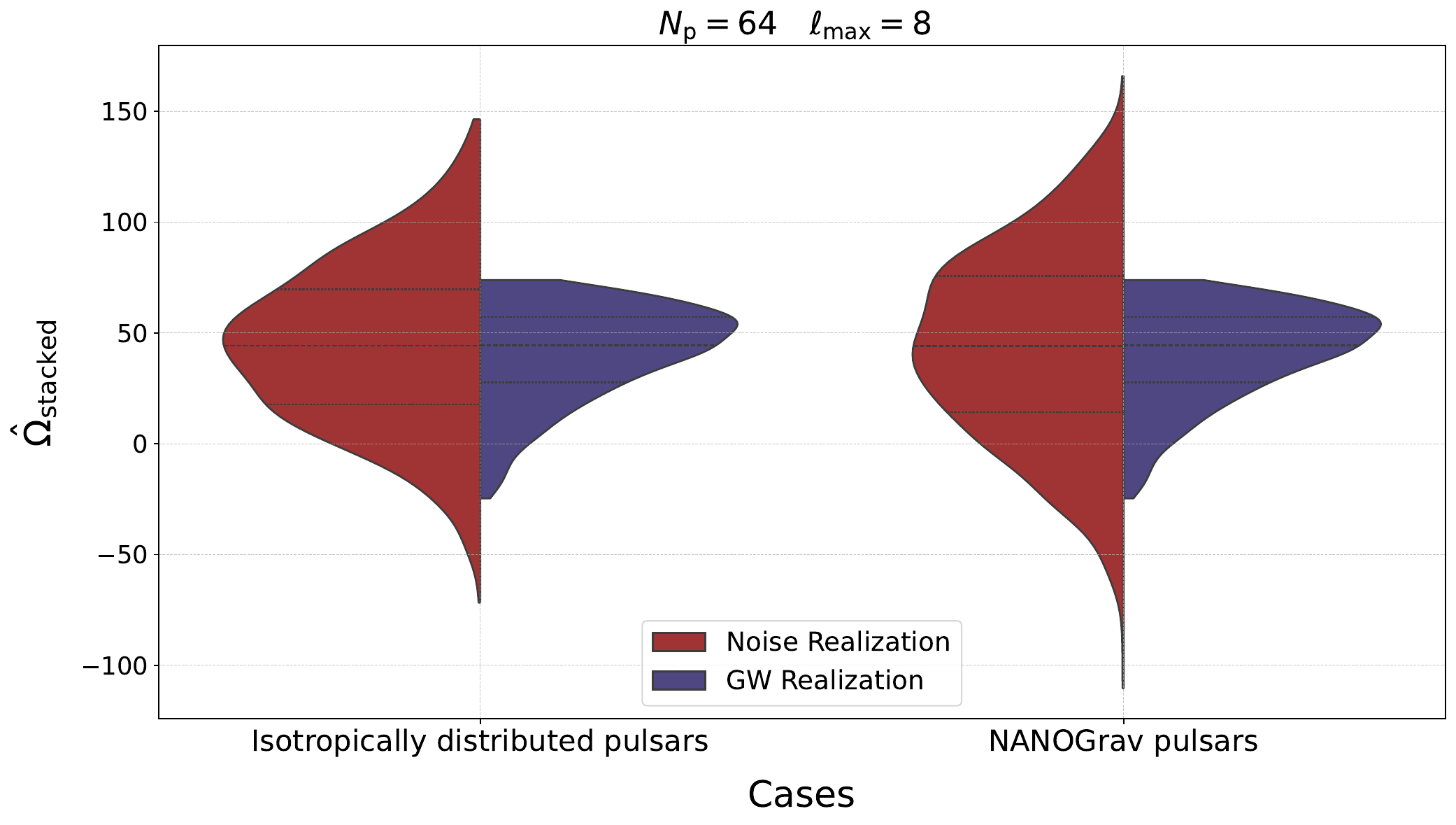}
    \caption{Violin plots of the stacked SGWB signal ($\hat{\Omega}_{\text{stacked}}$) for $\eta = 8.5$ and $\nu = 0$, comparing two scenarios: 64 isotropically distributed pulsars and 64 NANOGrav pulsars distributed anisotropically.}
    \label{fig:stack_aniso_pulsars}
\end{figure}

\begin{table}[ht]
\centering
\scalebox{1}{
\begin{tabular}{|c|c|c|c|c|c|}
    \hline
    \textbf{$N_{\rm p}$ $\ell_{\rm max}$} & \multicolumn{2}{c|}{\textbf{p-value (Frac = 0.5)}} & \multicolumn{2}{c|}{\textbf{p-value (Frac = 1)}} \\
    \cline{1-5}
    & \textbf{Noise Realization} & \textbf{GW Realization} & \textbf{Noise Realization} & \textbf{GW Realization} \\
    \hline
    $64 ~~ 8$    & 0.295 & 0.140 & 0.123 & 0.056 \\
    $100 ~~ 10$  & 0.224 & 0.133 & 0.047 & 0.048 \\
    $200 ~~ 14$  & 0.118 & 0.126 & 0.006 & 0.050 \\
    $400 ~~ 20$  & 0.051 & 0.121 & 0.001 & 0.041 \\
    $600 ~~ 24$  & 0.025 & 0.118 & 0.000 & 0.042 \\
    $800 ~~ 28$  & 0.014 & 0.108 & 0.000 & 0.040 \\
    \hline
\end{tabular}
}
\caption{We show the p-values of the $\hat{\Omega}_{\text{stacked}}$ distribution, defined as the fraction of the distribution less than or equal to zero, for different PTA resolutions. The results are shown for AGN fractions contributing to the SGWB of 0.5 and 1.}
\label{tab:table1}
\end{table}

\item\textit{\textbf{Impact of the anisotropic distribution of pulsars:}} 
The main impact of the anisotropic distribution of pulsars is that it results in a position-dependent SGWB noise map. However, this anisotropic noise does not affect the application of the stacking technique, as the method relies on identifying and summing the signal in pixels with high galaxy number density, rather than calculating any summary statistics like angular power spectrum.  In Fig. \ref{fig:stack_aniso_pulsars}, we present a violin plot comparing the distribution of stacking signals for two cases: one with 64 pulsars distributed isotropically and the other with the 64 NANOGrav pulsars \citep{nanogp}, with the same timing residual noise. The choice of 64 pulsars enables a direct comparison between an idealized isotropic distribution and a more realistic scenario, reflecting the approximate number of pulsars available in current datasets. The spatial anisotropy of the pulsar distribution introduces a corresponding anisotropy in noise, affecting the overall sensitivity of the measurement. In regions with a higher density of pulsars, the sensitivity is enhanced, whereas in regions with fewer pulsars, the sensitivity is significantly reduced. This non-uniform sensitivity alters the signal-to-noise ratio across the sky, leading to a net decrease in the overall sensitivity of the stacking signal. Consequently, the precision of the measured stacking signal is slightly degraded when pulsars are distributed anisotropically instead of isotropically. This underscores the impact of pulsar anisotropy on stacking signal measurements.

\item\textit{\textbf{Impact of redshift tomographic binning of AGN catalog:}} We want to determine whether the redshift bin that contributes most to the SGWB signal can be identified using the \texttt{Multi-Tracer Correlated Stacking} technique. To investigate this aspect, we perform stacking of $\Delta \Omega_{\rm gw}$ using AGN from specific redshifts range. In Fig. \ref{fig:stack_AGN}, we show the distributions of $\hat \Omega_{\rm stacked}$ as a function of AGN redshift bins, where two cases of supermassive BHB populations are considered: GW sources with a maximum redshift $z_{\rm max} = 3$ and GW sources with a maximum redshift $z_{\rm max} = 1$. Each of these cases is illustrated for Frac = 0.5 and Frac = 1. The stacking is performed using the AGN catalog restricted to five different redshift ranges, as specified on the x-axis. From the plot, we observe that the distributions of the stacked signal are consistent up to redshift bins below z = 3 for both cases. For the $z_{\text{max}} = 3$ scenario,  the stacking signal decreases gradually with increasing redshift. Similarly, for the $z_{\text{max}} = 1$ case, the stacking signal remains notably large up to z=3, despite the absence of any GW sources in this range for this scenario. Beyond z=3, the stacking signal decreases further but still retains a significant value.

This behavior suggests that the stacking method is not very sensitive to the tomographic distribution of the SGWB. While the technique successfully captures the overall correlation between galaxy distribution and SGWB anisotropy, it does not provide sufficient information to pinpoint the specific redshifts that dominate the contribution to the SGWB. In contrast, angular cross-correlation of the SGWB with galaxy distribution which explores the spatial clustering is an effective avenue to capture the correlation in redshift \citep{Sah:2024etc}. 
\end{enumerate}

\begin{figure}
  \centering
  \subfigure[]{\label{fig:stack_bin0.5}
    \includegraphics[width=0.8\linewidth,trim={0.cm 0  0 0.cm},clip]{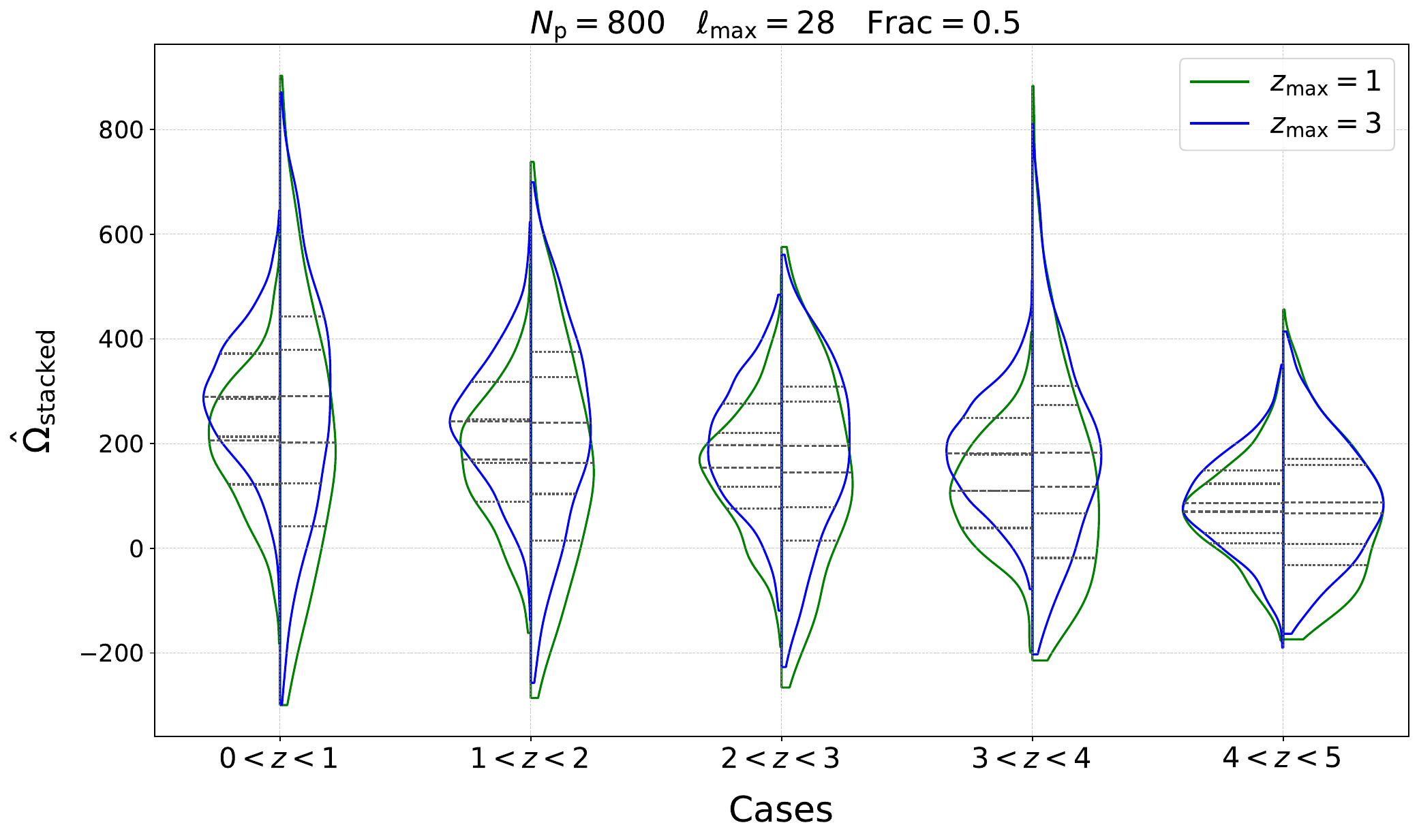}}
  \vspace{0.5cm}
  \subfigure[]{\label{fig:stack_bin1}
    \includegraphics[width=0.8\linewidth,trim={0.cm 0  0 0.cm},clip]{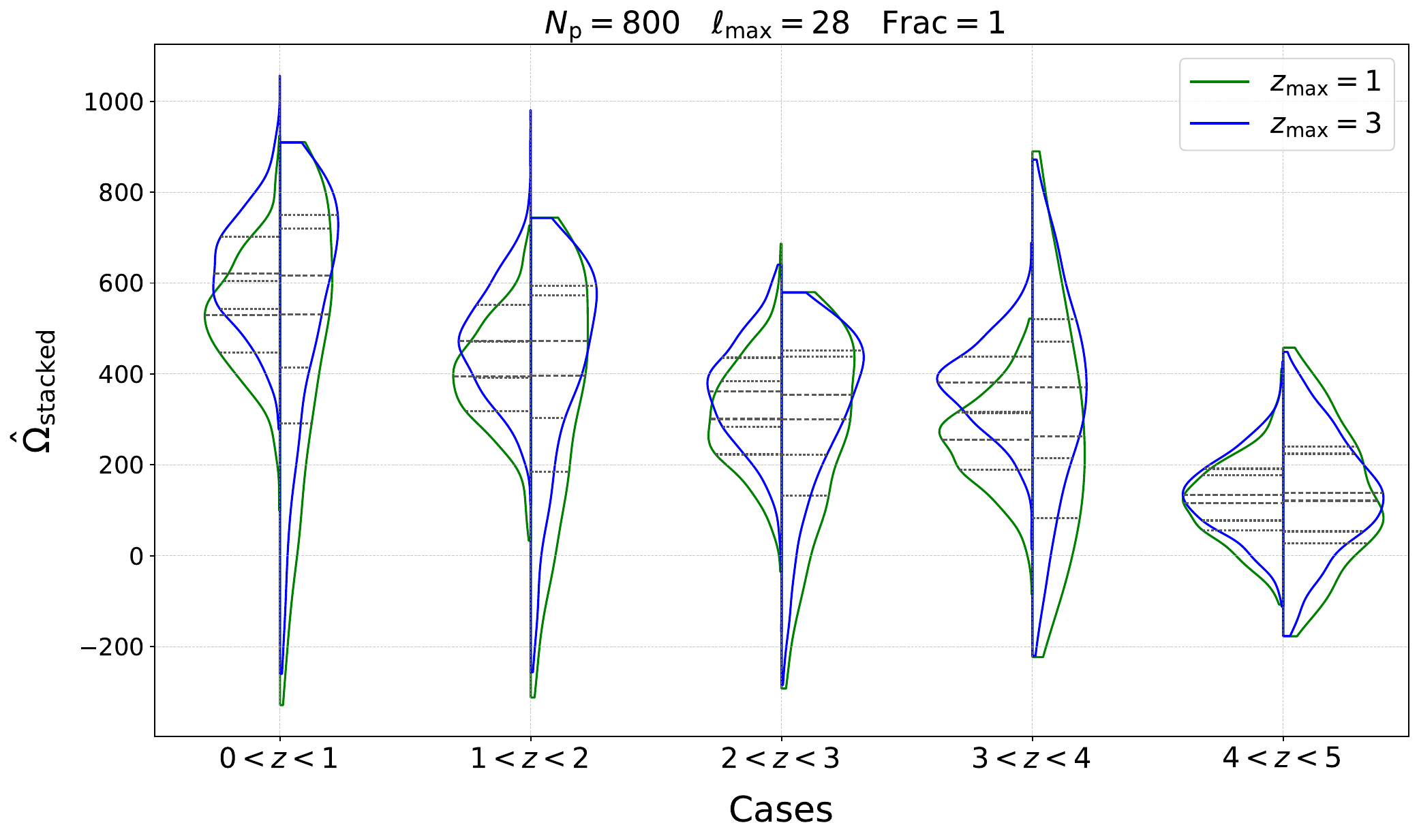}}
  \caption{Violin plots of the stacked SGWB signal ($\hat{\Omega}_{\text{stacked}}$) for two cases: sources with a maximum redshift of 3 (blue) and sources with a maximum redshift of 1 (green), stacked using different redshift bins of the AGN catalog. Panel (a) assumes an SGWB with an AGN fraction of 0.5, while panel (b) assumes an AGN fraction of 1. The left distribution in each plot represents variations induced by noise, and the right distribution reflects the impact of different GW realizations.}
  \label{fig:stack_AGN}
\end{figure}

\section{Conclusion and Future Prospects}\label{Conc}

In this paper, we introduced the \texttt{Multi-tracer Correlated Stacking} technique, a novel method designed to enhance the detection of anisotropies in the SGWB. By stacking the SGWB signal from regions of the sky with higher galaxy densities, we demonstrated that this approach is capable of detecting the anisotropic signal. The stacking technique effectively amplifies the anisotropic signal, offering a powerful tool for exploring the astrophysical origins of the SGWB. Furthermore, this technique can identify the underlying tracers of the SGWB signal and will help shed light on the formation and evolution of supermassive BHBs and their connection with different tracers.

Our analysis revealed that the stacking technique is capable of detecting the anisotropic and non-Gaussian signals originating from the supermassive BHB population. This method demonstrates a clear advantage over traditional approaches by its ability to differentiate between isotropically distributed GW sources and distribution that traces the galaxy distribution. Unlike conventional techniques such as angular power spectrum analysis, which are primarily capable of detecting only Gaussian anisotropic signal, the \texttt{Multi-tracer Correlated Stacking} technique can capture non-Gaussian signals as well. 
However, it is important to note a few limitations associated with the current demonstration. The effectiveness of the stacking method depends significantly on the sky coverage and the completeness of the tracers used. Limited redshift coverage or incomplete catalogs may lead to inaccuracies in identifying high-density regions, thereby reducing the efficiency of the technique.  Addressing these limitations will be critical for maximizing the potential of this method. In our future work, we will extend this to simultaneous inference using multi-tracers for catalogs which will be available from different large-scale structure surveys.

The application of the \texttt{Multi-tracer Correlated Stacking} technique on real data from upcoming surveys and experiments represents an exciting avenue for future work. As part of future work, we plan to apply this technique to observed data from current PTAs \citep{agazie2023nanograv,antoniadis2023second,zic2023parkes,xu2023searching,miles2024meerkat}. By leveraging the AGN catalog, we aim to analyze the observed SGWB maps for inferring anisotropies and test the alignment between the SGWB signal and large-scale galaxy structures. Though such measurements can be weak due to limited sky resolution (as shown in Fig. \ref{fig:stack_pul}), an interesting upper bound may be already possible from the current data.

This technique will be extremely promising in the future with the availability of data from the upcoming surveys. On one side using the Square Kilometre Array (SKA) \citep{maartens2015overview}, we can make precise timing of thousands of millisecond pulsars \citep{smits2009pulsar}. This will increase the spatial resolution and significantly improve the capability of the stacking technique. On the other side, future galaxy surveys such as the Legacy Survey of Space and Time (LSST) \citep{ivezic2019lsst} conducted by the Vera C. Rubin Observatory will provide deeper and more complete catalogs of galaxies, AGNs, and quasars. These surveys together will offer improved tracers for the supermassive BHB population, particularly at higher redshifts than the current catalogs, further boosting the precision of the stacking technique and discovering the connection between supermassive BHBs with their tracers using the anisotropic SGWB signal. In summary, the \texttt{Multi-tracer Correlated Stacking} technique will enable new discovery space in the area of astrophysics and cosmology in coming years using nHz GW signal. 

\section*{Acknowledgments}
This work is a part of the $\langle \texttt{data|theory}\rangle$ \texttt{Universe-Lab} which is supported by the TIFR and the Department of Atomic Energy, Government of India. The authors would like to thank the $\langle \texttt{data|theory}\rangle$ \texttt{Universe-Lab} for providing computing resources. The authors would also like to acknowledge the use of the following Python packages in this work: Numpy \citep{van2011numpy}, Scipy \citep{jones2001scipy}, Matplotlib \citep{hunter2007matplotlib}, Astropy \citep{robitaille2013astropy,price2018astropy}, Healpy \citep{gorski2005healpix,Zonca2019}, Seaborn \citep{Waskom2021} and Ray \citep{moritz2018ray}.

\bibliography{main}{}
\bibliographystyle{aasjournal}

\appendix
\section{Optimal Estimator of Stochastic Gravitational-Wave Background}
\label{sec:appendixA}

We use the optimal estimator of the Stochastic Gravitational-Wave Background (SGWB) $\Omega_{\rm gw}$ derived in  \cite{Anholm:2008wy} 
\begin{equation}
S_{IJ} = \int_{-\infty}^{\infty} df \tilde{s}_I^{*}(f) \tilde{s}_J(f) \tilde{Q}(f)
\end{equation}
where $\tilde{s}_I(f)$ and $\tilde{s}_J(f)$ are the Fourier transforms of the redshift of the signals from pulsars I and J respectively, and $\tilde{Q}(f')$ is the optimal filter 
\begin{equation}
\tilde{Q}(f) = \chi \frac{\overline{\Omega}_{\rm gw}(f) \Gamma_{IJ}}{f^3 P_I(f) P_J(f)}, 
\end{equation}
where $\Gamma_{IJ}$ is the overlap reduction function, $P_I(f)$ and $P_J(f)$ are the noise power spectra of the two pulsars and $\chi$ is the normalization constant choosen such that $<S_{IJ}>$ = $\overline{\Omega}_{\rm gw}(f_{ref}) \Gamma_{IJ}$. The uncertainty in the measurement of $<S_{IJ}>$ is given by
\begin{equation}
\sigma_{IJ} \sim \frac{\overline{\Omega}_{\rm gw}(f_{ref})}{\sqrt{T_{\rm obs}}} \left( \int_{-\infty}^{\infty} df \, \frac{|\overline{\Omega}_{\rm gw}(f)|^{2} }{f^6 P_1(f) P_2(f)} \right)^{-1/2}, 
\end{equation}
where $T_{\rm obs}$ is the total observation time. The overlap reduction function 
$\Gamma_{IJ}$ is given by
\begin{equation}
    \Gamma_{IJ} \propto \int  ~P(\hat{\omega})\Big(\mathcal{F}^{+}_{I}(\hat{\omega}) \mathcal{F}^{+}_{J}(\hat{\omega}) + \mathcal{F}^{\times}_{I}(\hat{\omega}) \mathcal{F}^{\times}_{J}(\hat{\omega})\Big), 
\end{equation}
where $\mathcal{F}^{A}_{I}(\hat{\omega})$ is the antenna response pattern of pulsar I, and $P(\hat{\omega})$ is given by
\begin{equation}
    P(\hat{\omega}) \equiv \frac{\Omega_{\rm gw}(f_{ref},\hat{\omega})}{\overline{\Omega}_{\rm gw}. (f_{ref})},    
\end{equation}
here $\Gamma_{IJ}$ can be expressed in an equal-pixel basis, also known as the radiometer basis \citep{mitra2008gravitational,thrane2009probing}. In terms of the sum over equal pixel areas $\Gamma_{IJ}$ can be represented as
\begin{equation}
    \Gamma_{IJ} \propto \sum\limits_{k}   ~P_k\Big(\mathcal{F}^{+}_{I,k} \mathcal{F}^{+}_{J,k} + \mathcal{F}^{\times}_{I,k} \mathcal{F}^{\times}_{J,k}\Big). 
\end{equation}
We can represent this in a matrix form as 
\begin{equation}
        \bm{\Gamma} = \textbf{R} ~ \textbf{P}, 
\end{equation}
where $\textbf{R}^{IJ}_{k}  \equiv \frac{3}{2} \frac{1}{N_{\rm pix}} ~ \sum\limits_A \mathcal{F}^{A}_{I,k}  \mathcal{F}^{A}_{J,k}  $ and $\textbf{P}_{k}$ $\equiv P_{k}$.

Using the above, we can write the likelihood function for the cross-correlations as \citep{pol2022forecasting} 
\begin{equation}
    \mathcal{P}(\hat{\textbf{S}}|\textbf{P}) \propto \textbf{exp}[\frac{-1}{2}(\hat{\textbf{S}} - \textbf{R} \textbf{P})^{T} \bm{\Sigma}^{-1} (\hat{\textbf{S}} - \textbf{R} \textbf{P})],
\end{equation}
where $\hat{\textbf{S}}_{IJ}$ = $S_{IJ}/\overline{\Omega}_{\rm gw}(f_{ref})$  is amplitude scaled optimal cross-correlation estimator, and  The $\Sigma$ is the amplitude scaled cross-correlation variance. $\Sigma$ is a diagonal matrix with diagonal elements given by $\Sigma_{IJ} = \Big(\sigma_{IJ}/\overline{\Omega}_{\rm gw}(f_{ref})\Big)^{2} = \bar{\sigma}_{IJ}^{2}$. The uncertainty in the measurement of $P_{k}$ denoted by $\mathbf{\Sigma}_{\rm pix}$ is given by the square root of the diagonal terms in the inverse of Fisher matrix  $\textbf{R}^{T} \bm{\Sigma}^{-1} \textbf{R}$ as 
\begin{equation}
    \begin{aligned}
        \mathbf{\Sigma}_{\rm pix} = &~  (\textbf{R}^{T} \bm{\Sigma}^{-1} \textbf{R})^{-1}. 
    \end{aligned}
\end{equation}
For an SGWB dominated by far-away point sources, the Fisher matrix can be approximated as a diagonal matrix, where its diagonal elements correspond to the diagonal elements of the original Fisher matrix \citep{romano2017detection}. In the present simulations, we find that about 90 \% of the SGWB is contributed by only $\sim 100$ brightest sources, justifying the applicability of the far-away point source approximation.

\end{document}